\documentclass{jfm}

\usepackage{graphicx}
\usepackage{natbib}
\usepackage{balu, ar, fancybox, setspace, rotating}

\usepackage{amsmath,amssymb,enumerate,acronym}
\usepackage{multirow}

\ifCUPmtlplainloaded \else
  \checkfont{eurm10}
  \iffontfound
    \IfFileExists{upmath.sty}
      {\typeout{^^JFound AMS Euler Roman fonts on the system,
                   using the 'upmath' package.^^J}%
       \usepackage{upmath}}
      {\typeout{^^JFound AMS Euler Roman fonts on the system, but you
                   dont seem to have the}%
       \typeout{'upmath' package installed. JFM.cls can take advantage
                 of these fonts,^^Jif you use 'upmath' package.^^J}%
      }
  \else
  \fi
\fi

\ifCUPmtlplainloaded \else
  \checkfont{msam10}
  \iffontfound
    \IfFileExists{amssymb.sty}
      {\typeout{^^JFound AMS Symbol fonts on the system, using the
                'amssymb' package.^^J}%
       \usepackage{amssymb}%
         \let\leq=\leqslant
         
      }{}
  \fi
\fi

\ifCUPmtlplainloaded \else
  \IfFileExists{amsbsy.sty}
    {\typeout{^^JFound the 'amsbsy' package on the system, using it.^^J}%
     \usepackage{amsbsy}}
    {\providecommand\boldsymbol[1]{\mbox{\boldmath $##1$}}}
\fi

\providecommand\bnabla{\boldsymbol{\nabla}}
\providecommand\bcdot{\boldsymbol{\cdot}}

\newsavebox{\astrutbox}
\sbox{\astrutbox}{\rule[-5pt]{0pt}{20pt}}

\title
{Nonlinear Evolution of a Baroclinic Wave and Imbalanced Dissipation}
\author[B.T. Nadiga]{Balasubramanya T. Nadiga$^1$%
  \thanks{Email address for correspondence: balu@lanl.gov}}
\affiliation{Los Alamos National Laboratory, MS-B214, Los Alamos, NM 87545, USA}

\begin{document}
\maketitle

\begin{abstract}
  We consider nonlinear evolution of an unstable baroclinic wave
  in a regime of rotating stratified flow that is of relevance to {\em
    interior} circulation in the oceans and in the atmosphere---a
  regime characterized by small large-scale Rossby and Froude numbers,
  a small vertical to horizontal aspect ratio, and no bounding
  horizontal surfaces. Using high-resolution simulations of the
  non-hydrostatic Boussinesq equations and companion integrations of
  the balanced quasi-geostrophic equations, we present evidence for a
  local route to dissipation of balanced energy directly through
  interior turbulent cascades. That is, analysis of simulations
  presented in this study suggest that a developing baroclinic
  instability can lead to secondary instabilities that can cascade a
  small fraction of the energy forward to unbalanced scales whereas
  the bulk of the energy is confined to large balanced scales.

  Mesoscale shear and strain resulting from the hydrostatic
  geostrophic baroclinic instability drive frontogenesis. The fronts
  in turn support ageostrophic secondary circulation and
  instabilities. These two processes acting together lead to a quick
  rise in dissipation rate which then reaches a peak and begins to
  fall slowly when frontogenesis slows down; eventually balanced and
  imbalanced modes decouple. 

  A measurement of the dissipation of balanced energy by imbalanced
  processes reveals that it scales exponentially with Rossby number of
  the base flow.  We expect that this scaling will hold more generally
  than for the specific setup we consider given the fundamental nature
  of the dynamics involved. 

  In other results, a) A break is seen in the total energy spectrum at
  small scales: While a steep $k^{-3}$ geostrophic scaling (where $k$
  is the three-dimensional wavenumber) is seen at intermediate scales,
  the smaller scales display a shallower $k^{-5/3}$ scaling,
  reminiscent of the atmospheric spectra of Nastrom \& Gage and b) At
  the higher of the Rossby numbers considered a minimum is seen in the
  vertical shear spectrum, reminiscent of similar spectra obtained
  using in-situ measurements.
\end{abstract}

\acresetall 

\section{Introduction}
Ocean circulation is mainly forced at large scales and instabilities
of the resulting large-scale circulation at scales related to the
internal Rossby radius of deformation give rise to intermediate-scale
mesoscale eddies \citep[e.g., see][]{vallis06, mcwilliams06,
  gill82}. The large-scale circulation and mesoscale eddies are
however in approximate geostrophic and hydrostatic momentum
balance. In this setting, classical geostrophic turbulence theory
suggests that such balanced dynamics leads to quasi-two-dimensional
behavior at the mesoscales and larger scales \citep{charney71}.  An
expected consequence of the quasi-two-dimensional behavior is an
inverse cascade of energy, wherein baroclinic instability injects
energy into the barotropic mode and nonlinear barotropic interactions
transfer barotropic (kinetic) energy to larger scales.  Although the
strength, extent, and nature of the barotropic inverse cascade itself
is an area of ongoing research \citep[e.g., see][and references
therein]{straub2014energy}, an established consequence of balanced turbulence
is a confinement of energy to large and intermediate scales. A side
effect of such confinement is to render small scale dissipation
ineffective in directly dissipating balanced mesoscale and large-scale
energy \citep[][and the extensive literature on geostrophic
turbulence]{charney71}
Thus a fundamental question regarding the energetics of ocean
circulation is as to how the ocean equilibrates in the presence of
continuous large-scale forcing and a tendency to confine energy to
large and intermediate scales.

The tendency of large-scale quasi-two dimensional (anisotropic)
rotating, stratified turbulence to confine energy to the large and
intermediate scales is in contrast to the forward energy cascade of
small-scale, three dimensional, isotropic turbulence \citep[e.g.,
see][]{frisch95, lesieur97}.  Here, by large-scales we mean the range
of scales down from the size of the domain to the first internal Rossby
radius of deformation. Elsewhere, we also refer to scales of the order
of the deformation radius as intermediate scales.  The scales and
phenomena that span these two asymptotic regimes of turbulence and the
interactions of the large-scale flow with boundaries are then expected
to hold the answer to the ocean-equilibration question above.  The
range of such processes include submesoscale and inertia-gravity wave
processes and they are further broadly categorized into boundary and
surface related processes, inertia-gravity wave processes, and other
instability processes \citep[e.g., see][]{molemaker05}. A large body
of recent research has focused on surface related processes in this
context. It is beyond the scope of the present article to review this
evolving field of research, and the reader is referred to, e.g.,
\citet{mcwilliams1998breakdown, molemaker05, thomas-tandon08,
  molemaker2010balanced, thomas2013symmetric} and references therein.
However, it is worth noting that there are a set of fluid dynamical
instabilities that underlie this line of research \cite[e.g.,
see][]{mcwilliams1998breakdown}.

Given that the atmospheric forcing of ocean circulation is mediated by
the top surface of the ocean, the importance of understanding the
dynamics of this region cannot be over-stated. However, and in
part for the same reason, the top surface is a preferred site for a
wide range of phenomena including, of course, thermal and momentum
exchange with the atmosphere. This leads to surface dynamics being
complicated. This renders the surface layer to not be the ideal
setting to study the instabilities that underlie
balanced-unbalanced interactions.

The aim of this article is two-fold. First, we propose and use an
idealized setup devoid of free surfaces---a periodic domain---to study
instabilities relevant for balanced-unbalanced interactions. Like in
the Charney model of baroclinic instability, baroclinic instability in
the setup we consider is due to interior gradients of potential
vorticity. In that sense the present setup is a version of the Charney
model where the base state is periodic (and presently
$\beta=0$). Equivalently, the setup may be thought of as a continuously
stratified, periodic version of the Phillips model. There is a
further advantage to the present setup: In the Charney model, the
zonal thermal wind of the base state is independent of the meridional
coordinate and depth.  Clearly the advantage of this assumption (and
the constant scale-height assumption that lead to meridional gradient
of potential vorticity being constant) is in reducing complexity and
giving insights into the dynamics of baroclinic instability. However,
in more realistic settings, baroclinicity is not uniform; it is more
typical for it to be concentrated in zonal bands, as for example in
the Antarctic Circumpolar Current or the jet-streams. The present
setup, by generalizing the meridional and depth dependence of the
zonal thermal wind to be periodic can represent the more typical
non-uniform distribution of baroclinicity.

More fundamentally, of course, the question remains as to whether
surface- and boundary-related processes are the only pathways that
lead to a forward cascade of balanced energy.  Indeed, e.g., as
pointed out in Molemaker et al., 2005\nocite{molemaker05}, the bulk of
the balanced energy resides in the vertical interior and there is the
alternative (spatially) more local route to dissipation directly
through the interior turbulent cascade. This will be the second aim of
the article---to explore the interior route to dissipation of balanced
energy. We will do this by studying an initial value problem in which
practically all of the energy in the initial condition resides at
a scale close to the system size. But this initial condition is chosen
to be baroclinically unstable, so that the occurrence of baroclinic
instability initiates transfer of energy to smaller scales, primarily
of course to the mesoscales. The resulting mesoscale shear and strain
can then drive frontogenesis and the resulting frontal regions are
susceptible to other unbalanced instabilities. This chain of events
thus has the potential for establishing an interior pathway to
dissipate balanced mesoscale energy. An aim of this article is to
study this pathway, quantify the resulting small-scale dissipation,
and establish its dependence on the Rossby number of the base flow.

\section{The Small Rossby number and Froude number Regime}
In this article, our interest is in a regime that is of relevance to
ocean circulation (and to a certain extent to atmospheric
circulation). In particular a) the horizontal scale of the domain
($L$) we are interested in corresponds to that of the oceanic
mesoscales in the mid-latitudes ($\approx 100$ km corresponding to a
few times the first internal Rossby deformation radius). The average
depth of the ocean ($H$) is about 5km leading to a (vertical to
horizontal) domain aspect ratio $H/L$ of about 0.05, with this number
being smaller yet if only the upper ocean were to be considered. At
these scales, we are interested in b) the velocity-based Rossby number
defined as $\Ro = U/fL$ where $U$ is peak velocity is small (say, $<
0.1$), and c) the velocity based (vertical) Froude number defined as
$\Fr = U/N_0H$ is small (say, $< 0.1$). For a comprehensive overview
of the analytical treatment of the Boussinesq equations in this
regime, see \citet{babin1997asymptotic}. We note that since the
nondimensional form of the rotating-stratified Boussinesq equations
(see next section) have three non-dimensional parameters (e.g., Rossby
number \Ro, Froude number \Fr, and vertical to horizontal aspect ratio
\AR), it is important that they all be considered when making
comparisons of dynamic similarity to other studies. For example, we
note that there are a number of studies that consider small $\Ro$ and
$\Fr$ in a unit aspect ratio domain \citep[e.g., see][]{bartello95,
  smith-waleffe02}.  Without getting into details, we suspect that
vertical velocities in these flows tend to be higher than what is
typical of oceanic mesoscales and submesoscales.

The smallness of the Rossby and Froude numbers of the initial
condition we choose would be considered such as to lead to a balanced
nonlinear flow evolution. One argument in favor of this point of view
is as follows: The tendency of balanced quasi-two-dimensional
turbulence to confine energy to the large and intermediate scales is
accompanied by a forward cascade of potential enstrophy at small
scales. The theoretical spectral scaling of the energy spectrum in
this second inertial range corresponds to a steep $-3$ slope
\citep{charney71}. Such $-3$ (and steeper slopes due to the presence
of coherent structures and other modeling artifacts) have repeatedly
been verified numerically in both the quasi-geostrophic approximation
\citep[e.g., see][]{lesieur97} and the more complete non-hydrostatic
Boussinesq description \citep[e.g., see][]{bartello95}. A consequence
of such a steep spectral fall-off of the energy spectrum is that it
precludes the possibility of increasing (appropriately defined scale
dependent) Rossby number as the scale decreases. Thus, if Rossby
number could not increase with decreasing scale, there would not be a
route for initially balanced, small Rossby and Froude number flows to
break out of the scaling and become imbalanced. In the absence of
external forcing and surface-related processes, it may therefore be
tempting to think of this argument as a self-consistent explanation
for why such flows should remain balanced.

However, there is the possibility that these arguments can fail in the
context of the non-hydrostatic Boussinesq equations. A couple of
heuristic lines of reasoning for why the above arguments can fail are
as follows: It is possible that the nonlinear term in the definition
of potential vorticity (see equations \ref{eq:pv} and \ref{eq:ndpv})
could become increasingly important as resolution is increased. In
such a situation, potential enstrophy would no more be a quadratic
quantity, leaving total energy as the single (inviscidly) conserved
quadratic (see section \ref{sec:energy}) invariant and opening up the
possibility of a forward cascade of energy, a situation somewhat
similar to three-dimensional turbulence.  

From a more dynamical point
of view, Hoskins et al., 1978 \nocite{hoskins1978new} point out that
in quasi-geostrophic theory, the role of {\em ageostrophic} motion is
to restore thermal wind (geostrophic and hydrostatic) balance which
the geostrophic motion is tending to destroy by changing the two parts
of the thermal wind balance equally but in opposite directions. In
this sense, the exact cancellation of the two tendencies by the
ageostrophic velocity is more an artifact of the construction of the
quasi-geostrophic approximation than it is a reflection of
reality. Consequently, the ageostrophic velocity in the Boussinesq
system could be such that it does not enforce exact cancellation and
thus lead eventually to imbalanced motions. 

Quasigeostrophy-related $-3$ (and steeper) scaling of the energy
spectrum in the (full) enstrophy cascade range has been obtained in
simulations of the Boussinesq equations
\citep[e.g.][]{waite2006transition}. While shallower spectral slopes
have been observed at the small-scale end of the enstrophy cascade
range in other numerical simulations \citep[e.g.][]{bartello95,
  waite2006transition}, they have typically been attributed to the
large-scale Rossby number being large.  (Note that the scale at which
the break in slope appears is likely dependent on the choice of
initial condition in the initial value problem and on the choice of
forcing in the forced setting.)  

That a shallow small-scale tail is
always possible irrespective of how small the large-scale Rossby
number is, is a possibility that has not been sufficiently considered.
Indeed, we find that the scale at which the shallower slopes are seen
decreases continuously when the large-scale Rossby number is reduced
(see Fig.~\ref{fg:te_spec}). This finding supports the above-mentioned
possiblity of a shallow tail at any large-scale Rossby number and at
the same time highlights the high resolutions that are required to
capture imbalanced motions when those Rossby numbers are small.

At this point, we digress to point out another difference in our
experimental design. Previous studies of the initial value problem in
the strongly rotating and strongly stratified flow regime that use the
full nonhydrostatic Boussinesq equations have focussed mainly on
statistics of the initial condition such as scale distribution of
energy \citep[e.g.,][and others]{bartello95, smith-waleffe02, waite2006transition} . While we consider such statistics as well, our initial
conditions are chosen to represent the uni-directional thermal-wind
balance that is common place in both the atmosphere and the world
oceans. Further, the baroclinic instability of such a flow is
fundamental to atmospheric and oceanic circulation and it is the
imbalance triggered by this instability that we consider. 

The above discussion related to the nonlinear evolution of initially
balanced flows implicitly assumed that imbalance is easily
identified. However, this may not always be the case even though
imbalance is associated with zero potential-vorticity modes. A simple
way of recognizing the difficulty in identifying imbalance is to note
the fact that balance can exist at different orders. For
example, optimal potential vorticity balance of
\citealp{viudez2004optimal} or that of
\citealp{mcwilliams1985uniformly} are balances at an order higher than that
associated with thermal-wind. Thus the definition of balance can be
complicated, there exists a substantial body of literature dedicated
to it \citep[e.g. see][and references therein]{viudez2004optimal}, and
it is an ongoing area of research. For this reason, we will take a
pragmatic approach to imbalance in this article (while hoping to
follow up on this issue in the future): Rather than diagnose imbalance
from a dynamics perspective, we will use an energy perspective. This
is because, from the point of view of energy in rotating stratified
flows, it seems sufficient to consider those aspects of dynamics of a
flow as balanced that tend to confine energy to large and intermediate
scales. Other aspects of dynamics that lead to a forward cascade of
energy can then be considered unbalanced.

As such, we will monitor various aspects of the flow to examine the
possibility of forward cascade of energy over the course of nonlinear
evolution of the baroclinically-unstable initial condition. At
the lowest order, we will monitor domain-integrated dissipation of
energy. Once robust behavior is seen in dissipation, we
 will examine spectral slopes. In the analysis of spectral slopes, the appearance of
slopes  shallower that $-3$  at the smaller of the resolved scales
will be taken as an indicator of forward cascade resulting from
unbalanced processes.  In this sense, this study shares similarities
with various previous studies in the atmospheric circulation modeling
literature that attempt to explain the spectral slopes observed in
Nastrom and Gage, 1985\nocite{nastrom85}. Finally, and to a lesser
extent we will examine the spectral flux of energy across scales using
the framework of (\ref{eq:transfer}) to further validate our findings.

The rest of the paper is organized as follows. The next section
describes the formulation of the problem. This is followed by a
section on numerical simulations where nonlinear evolution of a
particular flow configuration is analyzed in detail under the
quasi-geostrophic approximation and under the non-hydrostatic
Boussinesq approximation and at different resolutions. Evidence of
imbalance is presented in this section. This is followed by a section
where the scaling of dissipation with Rossby number is
considered. After considering spectral scaling and spectral fluxes in
the following section, we summarize and conclude with some discussion.

\section{Problem Formulation}
We consider the nonlinear evolution of a zonal flow
that is in geostrophic and hydrostatic (thermal wind) balance in a
stably-stratified environment 
$$\rho(x,y,z,t) = \rho_0 + \rho_{env}(z) + \rho'(x,y,z,t)$$
and in a uniformly rotating Cartesian frame.
In order to focus on the role of interior processes, boundaries are
explicitly eliminated by considering the above evolution in a
triply-periodic domain. The latter feature of the model setup is in
common with numerous previous studies such as those of Bartello, 1995;
Smith \& Waleffe, 2002; Bartello, 2010 and others. We consider
evolution under both the non-hydrostatic Boussinesq system that allows
for imbalanced processes and the quasi-geostrophic system that allows
for only balanced dynamics.

\subsection{The Non-hydrostatic Boussinesq System}
We write the non-hydrostatic Boussinesq equations
in the nondimensional form as:
\begin{gather}
{\partial \vv{u}_h \over \partial t} + \vv{u} \bcdot \bnabla \vv{u}_h
+  \frac{\vv{k} \times \vv{u}_h}{\Ro} = -\frac{\bnabla_{\!h} \phi}{\Ro}
+ D_{\vv u_h}
\cr
\AR^2
\left({\partial w \over \partial t} +
  \vv{u} \bcdot \bnabla w \right) =  
-\frac{1}{ \Ro} \frac{\partial \phi}{\partial z} + \frac{b}{\Fr} 
+ D_w
\cr
{\partial b \over \partial t} + \vv{u} \bcdot \bnabla b +  \frac{w}{\Fr}
=  
+ D_b
\cr
\bnabla\bcdot\vv{u} = 0.
\label{eq:bouss}
\end{gather}
Here, beyond the usual notation, subscript $h$ corresponds to
horizontal ($\vv{u}_h$ is the velocity vector in the
horizontal---(x,y)---plane, and $\bnabla_h$ is the horizontal gradient
operator), $w$ is the vertical velocity, $b$ is 
buoyancy ($b = -\frac{g}{\rho_0} \rho',$)
and where we have used $D$ to represent the dissipation operator for
brevity. The dissipation operator itself is discussed further later in
this section.

The scales chosen for the non-dimensionalization above are such as to
highlight the geostrophic and hydrostatic balance that is typical of
large scale flow in the ocean and atmosphere and the small aspect
ratio of the domains in which such flow occurs. In particular,
horizontal lengths are non-dimenionalized by $L$, the the horizontal
length of the domain (square in the x-y directions), vertical lengths
by $H$, the depth of the domain (z direction), horizontal velocities
by $U$ the peak velocity of the initial zonal flow, time by $L/U$,
vertical velocities by $UH/L$ (cf. divergence free condition), dynamic
pressure by $\rho_0 f U L$ (cf. geostrophy) where $f$ is the Coriolis
frequency on the f-plane we consider, and perturbation buoyancy by $U
N_0$ (in quasi-geostrophy, $b=f\partial \psi/\partial z; N_0H \approx fL$)
where $N_0$ is the Brunt-Vaisala frequency given by 
\begin{gather}
N_0 = \sqrt{-\frac{g}{\rho_0} \frac{d\rho_{env}}{dz}}.
\end{gather}
Finally we note that $U$ is the scale of the geostrophic velocity in
balance with buoyancy $b$.

\subsection{The Quasi-Geostrophic Approximation}
\label{sec:qg}
The above Boussinesq system conserves a potential 
vorticity \citep[e.g., see][]{bartello95} given by 
\begin{gather}
Q = \bomega_a \bcdot \bnabla b = 
(\bomega + f \widehat{\vv{z}}) \bcdot \left(N_0^2 \widehat {\vv z} + \bnabla b\right) = 
f N_0^2 + 
\left(\zeta N_0^2 + f \frac{\partial b}{\partial z}\right) + 
\bomega \bcdot \bnabla b,
\label{eq:pv}
\end{gather}
where $\bomega$ is (relative) vorticity $\bomega_a$ is absolute
vorticity, $\zeta$ is the vertical component of relative vorticity and
the first, second and third terms are respectively independent, linear
and quadratic in velocity and buoyancy variables. 
The nondimensional form is given by 
\begin{gather}
Q = 1 + \Ro\zeta +  \Fr\frac{\partial b}{\partial z} +
\Ro \Fr \, \bomega \bcdot \bnabla b,
\label{eq:ndpv}
\end{gather}
where all variables are now nondimensional.  However, (\ref{eq:pv}) or
(\ref{eq:ndpv}) cannot be inverted to obtain {\em all} dynamical
quantities of interest since there are other modes---those with zero
potential vorticity---that evolve as well.  Therefore, inverting
potential vorticity or an approximation of it (e.g., its
linearization) is useful only to the extent that it can help diagnose
the balanced component.  

In the low Rossby number, low Froude number
regime that we are concerned with in this article, the large-scale,
low-frequency modes of the system are well described by the
quasi-geostrophic approximation.  This system is obtained by making
the hydrostatic approximation ($\AR\ll 1$) and performing asymptotic
analysis of the above equations using $\Ro$ and $\Fr$ as small
parameters \citep[e.g., see][]{vallis06}; the quasi-geostrophic
approximation is obtained at the lowest order in $\Ro$ and $\Fr$.  In
this approximation, advection is reduced to that by the horizontally
non-divergent horizontal geostrophic velocity and buoyancy is related
to the vertical derivative of the geostrophic streamfunction as
$$b = \frac{\Fr}{\Ro}\,\frac{\partial \psi}{\partial z}$$
(dimensionally $b = f\frac{\partial \psi}{\partial z},$).
In the inviscid adiabatic limit, the
buoyancy equation reduces to 
\begin{gather}
{\partial b \over \partial t} + \vv{u}_g \bcdot \bnabla_h b = -
\frac{w}{\Fr},
\label{eq:qg_buoy}
\end{gather}
and the curl of the horizontal momentum equation reduces to 
\begin{gather}
{\partial \zeta \over \partial t} + \vv{u}_g \bcdot \bnabla_h \zeta = -
\frac{1}{\Ro}\frac{\partial w}{\partial z}.
\label{eq:qg_rv}
\end{gather}

On eliminating $w$ from (\ref{eq:qg_rv}) and (\ref{eq:qg_buoy}),
evolution of the system can be described in terms of a single scalar,
the quasi-geostrophic potential vorticity:
\begin{eqnarray}
\frac{\partial q}{\partial t} + \vv u_h \bcdot \bnabla_h q = 0\cr
q = \bnabla_h^2 \psi + \frac{\Fr^2}{\Ro^2} \frac{\partial^2
  \psi}{\partial z^2}
\label{eq:qg}
\end{eqnarray}
where an f-plane and constant stratification have been assumed. In
this case, the quasi-geostrophic potential vorticity can be inverted
to obtain all relevant quantities in the approximation. In particular
once the scaled-Laplacian is inverted to obtain $\psi$, the
streamfunction, $\vv u_h = \widehat {\vv z} \times \psi$. Finally,
comparing the expressions for potential vorticity in (\ref{eq:qg}) and 
(\ref{eq:ndpv}), the quasi-geostrophic potential vorticity is seen to be
the linear component of potential vorticity that is conserved in the
Boussinesq system. (The absence of the constant term in
  (\ref{eq:qg}) leads it to be qualified further as perturbation
  quasi-geostrophic potential vorticity.)

\subsection{Energy Equation}\label{sec:energy}
Kinetic energy (per unit mass) is defined in the usual fashion as
$$k = \frac{1}{2}\left(\vv u_h^2 + \AR^2 w^2\right),$$
and the available potential energy (per unit mass) as
$$p = \frac{b^2}{2},$$ 
where the quantities are averaged over the volume of interest and noting that the
variables are non-dimensional. (In the dimensional form $p_d =
b_d^2/(2N_0^2)$ where the subscript $d$ stands for ``dimensional''.) 

The equations for the evolution of kinetic energy and potential
energy are easily formed starting from  (\ref{eq:bouss}) by taking the
dot product of the momentum equation with velocity and multiplying the
buoyancy equation by the perturbation buoyancy. They may be
found in textbooks \citep[e.g., see][]{vallis06} and since we do not
require the details of those equations, we will not reproduce them
here.  However, we note that integrating the kinetic and potential
energy equations over the domain and summing the two equations  leads to
$$ \frac{d T}{dt} = -D,$$
where $T = K + P$ is the domain-integrated total energy, $K$ and $P$
domain-integrated kinetic and potential energy respectively and $D$ is
the domain-integrated dissipation of total energy by small scale
dissipation.

In order to better understand dynamics including that of dissipation,
besides studying the evolution of domain-integrated energy, we will
also consider the flux of energy across scales. For
this purpose, we take the dot product of the Fourier transform of the
momentum (buoyancy) equation in (\ref{eq:bouss}) and  the conjugate
of the Fourier transform of velocity (buoyancy) to arrive at equations
that describe the evolution of energy in a scale-wise fashion:
\begin{gather}
\frac{d \sK(\bkappa)}{d t} + \sK_\str(\bkappa) = {\cal P\!W}_\str +
{\cal B\!F}_\str + \sD_\sK(\bkappa)\cr
\frac{d \sP(\bkappa)}{d t} + \sP_\str(\bkappa) = 
- {\cal B\!F}_\str + \sD_\sP(\bkappa)\cr
\frac{d \sT(\bkappa)}{d t} + \sT_\str(\bkappa) = {\cal P\!W}_\str +
\sD_\sT(\bkappa),
\label{eq:transfer}
\end{gather}
where 
$\sT_\str(\bkappa) = \sK_\str(\bkappa) +
\sP_\str(\bkappa)$ is the nonlinear scale transfer of total energy,
with $\sK_\str(\bkappa)$, $\sP_\str(\bkappa)$ representing the
nonlinear scale transfer of kinetic and potential energy respectively, and
$\sD_\sT(\bkappa) = \sD_\sK(\bkappa) + \sD_\sP(\bkappa) $ is 
scale-wise dissipation. In other notation, ${\cal P\!W}_\str$ stands
for the scale transfer of kinetic energy by pressure work and ${\cal
  B\!F}_\str$ represents the scale transfer of kinetic/potential
energy by buoyancy flux. 

For convenience, we consider only the
shell-averaged version of this equation (one-dimensional form) and
prefer the nonlinear flux 
of energy across scales $\sT_f$, $\sK_f$, and $\sP_f$---obtained as integrals of the respective transfer as 
$$\sT_f(k) = \int_0^k \sT_\str dk.$$

\subsection{Initial Conditions}
\begin{figure}
\centering
\includegraphics[trim=0 0 0 0, clip, width=0.49\textwidth]{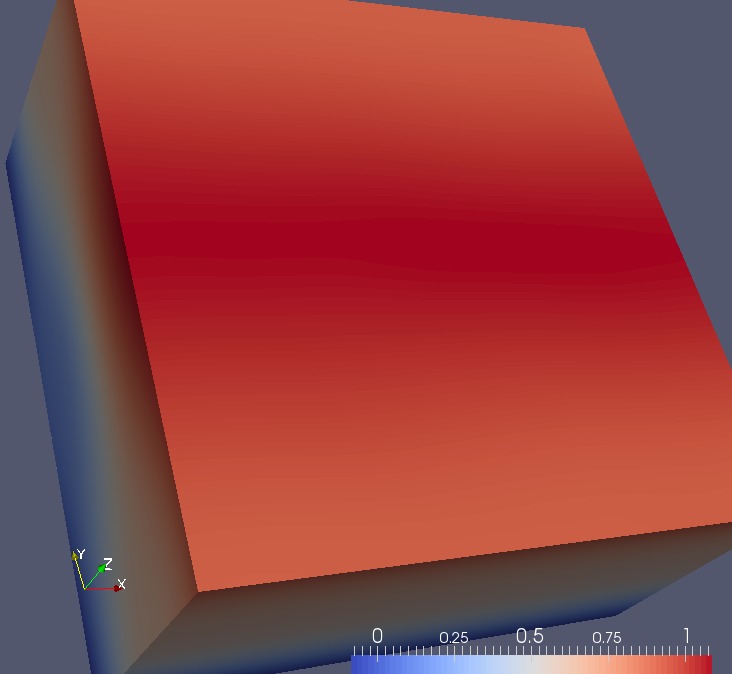}
\includegraphics[trim=0 0 0 0, clip, width=0.49\textwidth]{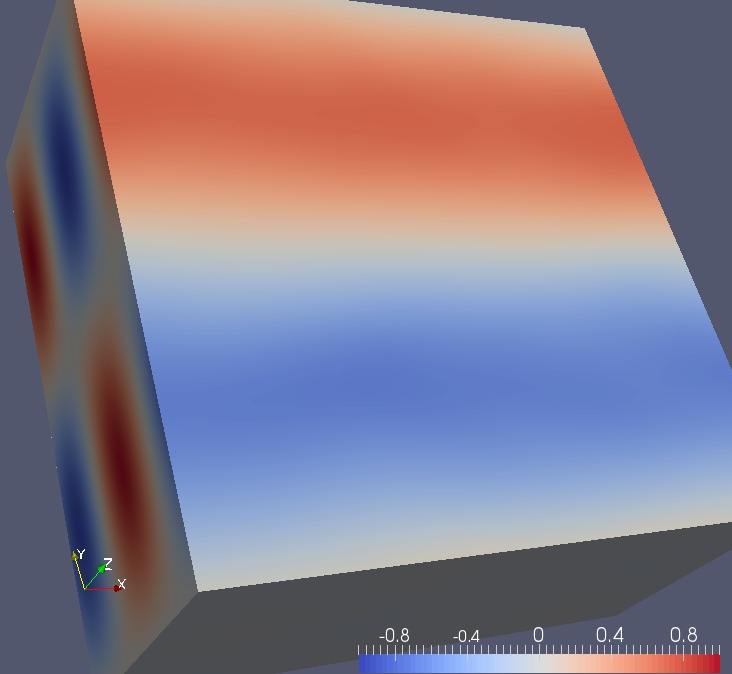}
\includegraphics[trim=0 0 0 0, clip,
width=0.49\textwidth]{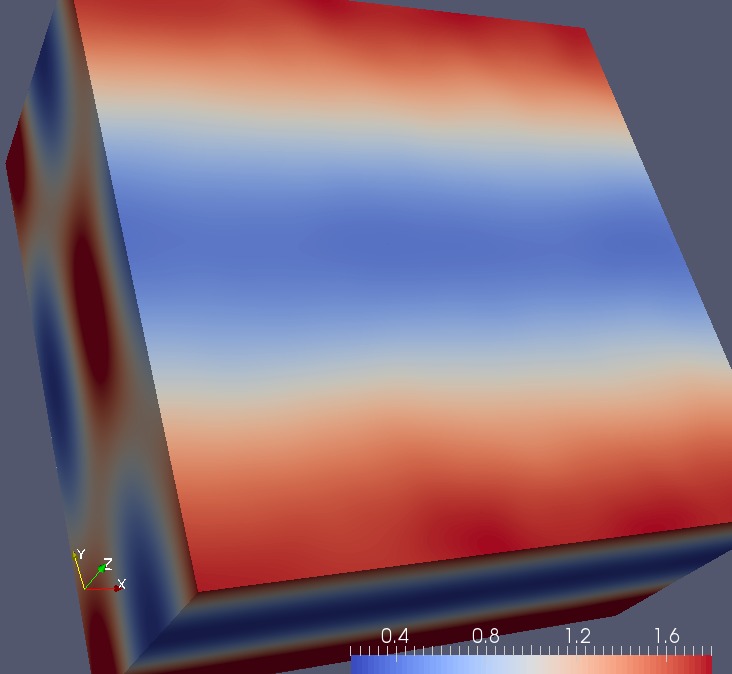}
\includegraphics[trim=0 0 0 0, clip,
width=0.49\textwidth]{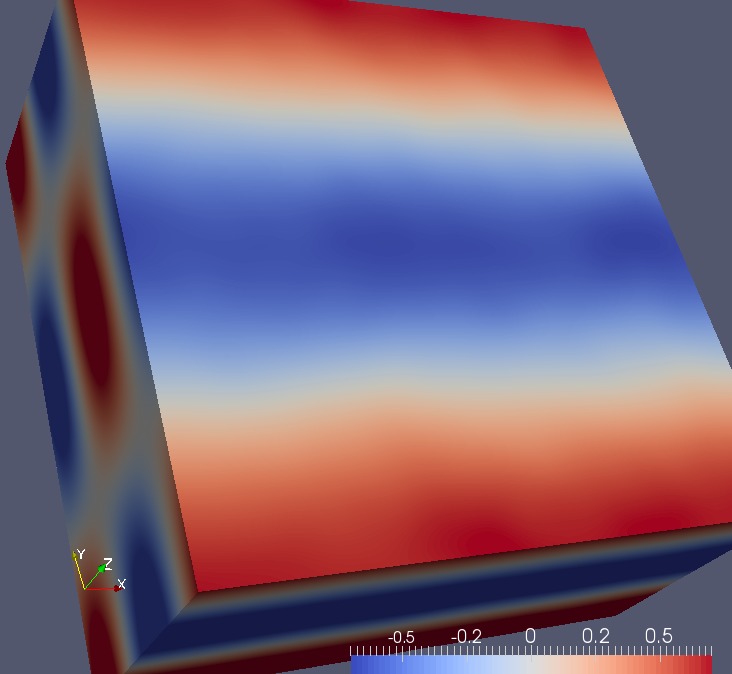}
\caption{Initial conditions: Buoyancy (top left), zonal velocity (top
  right), and potential vorticity (bottom left) for the Boussinesq
  system. Perturbation potential vorticity for the QG system (bottom
  right).The vertical has been exaggerated by a factor of 10. The
  actual vertical extent is a twentieth of the horizontal extent. }
\label{fg:IC}
\end{figure}

We choose the initial condition to be in geostrophic thermal-wind
balance:
\begin{gather}
{\partial \over \partial z} \left(v, -u\right) = {\Ro \over
\Fr}\bnabla_h b, 
\; w=0
\label{eq:thermalwind}
\end{gather}
where, as before $(u,v,b)$ represents a perturbation about the stably
stratified no-flow hydrostatic equilibrium.  A particular form of the
perturbation $b$ that further satisfies periodicity is of the form:
\begin{gather}
b(x, y, z) = cos(2\pi k_x x) cos(2\pi k_y y) cos(2\pi k_z z),
\label{eq:taylorgreen}
\end{gather}
somewhat reminiscent of the Taylor-Green vortex solution, but now in
the extended velocity and buoyancy variables, and with velocity given
by (\ref{eq:thermalwind}) \citep[also see][]{simon13}. In
(\ref{eq:taylorgreen}), $k_x$, $k_y$, and $k_z$ are the
non-dimensional wavenumbers in the three spatial directions. After
noting that uni-directional flows in this set are exact solutions of
the inviscid adiabatic equations, we choose the gravest
uni-directional (horizontal) flow with no variations in the flow
direction as the base state.  In particular, we choose the flow to be
zonal and with no variations in the zonal ($x$) direction ($k_x=0,
k_y=1, k_z=1$):
\begin{gather}
b(x, y, z; t=0) = cos(2\pi k_y y) cos(2\pi k_z z),
\label{eq:base}
\end{gather}
The initial buoyancy ($-g (\rho_{env} +
  \rho')/\rho_0$ is plotted in the nondimensional form as $z + \Fr\,
  b$.), zonal velocity, and potential vorticity fields are shown in
Fig.~\ref{fg:IC}. These fields are for the specific Rossby and Froude
numbers to be considered in the next section. For the other Rossby and
Froude numbers considered in this article, it should be clear that the
buoyancy and zonal velocity fields are simply scaled by the respective
Froude and Rossby numbers. Further, for the other Rossby and Froude
numbers, the structure of the potential vorticity field remains
visually indistinguishable. This is because, and as discussed further,
over the range of Rossby and Froude numbers considered, the quadratic
term in the definition of potential vorticity is small.

It is easy to see that the quasi-geostrophic potential vorticity of
the unidirectional base flow is proportional to $cos(2\pi y) sin(2\pi
z)$. This can also be seen in the right panel of Fig.~\ref{fg:IC}.
This form satisfies the Charney-Stern-Pedlosky necessary condition for
baroclinic instability which requires that $\partial q/\partial y$
change sign in the domain \citep[e.g., see][]{vallis06}. 

For the Boussinesq system, the linear term in (\ref{eq:ndpv}) is
likewise proportional to $cos(2\pi y) sin(2\pi z)$ as seen in the
expression for potential
vorticity of the base state in (\ref{eq:nbpvic}):
\begin{gather}
  Q(x,y,z;t=0) = 1 + \Ro \left(1 + \left(\frac{\Fr}{\Ro}\right)^2 \right)
  \frac{\partial b}{\partial z} + \Ro^2\left(\frac{\Fr}{\Ro}\right)^2\left[
    \left( \frac{\partial b}{\partial z} \right)^2 - \left(
      \frac{\partial b}{\partial y} \right)^2 \right].
\label{eq:nbpvic}
\end{gather}
For the Boussinesq system,
initial conditions considered in this article are such that $\Fr/\Ro$,
the ratio of the domain size to Rossby radius, is held fixed.
Consequently, for these cases, the ratio
of the magnitude of the quadratic term to the magnitude of the linear
term ($1.6\pi\Ro$) is between 0.10 and 0.17. 

Although the initial conditions satisfy the necessary condition for
instability, integration of the unidirectional flow over several tens
of eddy turnover times did not lead to a destabilization of the base
flow solution. That is to say, the truncation errors were too small to
destabilize the flow in a reasonable length of time.  Therefore, a
further set of uni-directional thermal-wind modes that are again
individually exact solutions were superposed on the base flow.  These
modes were chosen to have an energy spectrum that fell off steeply as
$(k_hk_z)^{-8}$ (essentially confined to the largest scales), and to
contain a small fraction of the energy of the unidirectional base flow
(approximately $0.05\%$); these perturbations cannot be seen in
Fig.~\ref{fg:IC}).

\subsection{Numerical Discretization and Consistency Between
  Horizontal and Vertical Resolutions}
A standard fully-dealiased Fourier pseudo-spectral method is used for
the spatial discretization and the equations are integrated in time
using a fourth order Runge-Kutta scheme. The above discretization, as
implemented in parallel using MPI in the Sandia-LANL DNS code
\citep{taylor2003recovering} forms the computational kernel used 
for the simulations. 

We next consider the issue of consistency between horizontal and
vertical resolutions. In quasi-geostrophic dynamics horizontal scales
$\Delta L$ and vertical scales $\Delta z$ are related as \cite[e.g.,
see][]{lindzen1989consistent}
$$\frac{\Delta L}{\Delta z} = \frac{N_0}{f},$$
where $\Delta L$ is related to the Rossby radius. It is clear that the
quasi-geostrophic approximation is applicable to horizontal scales of
the order of the first Rossby radius. However, we'll presently be
considering quasi-geostrophic simulations simply as balanced
counterparts (controls) of the unbalanced Boussinesq simulations, and
for this reason, the quasi-geostrophic simulations will consider
horizontal scales much smaller than the first Rossby radius. Thus if the
smallest Rossby radius that is resolved is of the order of the grid
resolution $\Delta x$, then to adequately resolve that mode in the
vertical, the vertical resolution has to satisfy
\begin{equation}
\Delta z = \frac{f}{N_0} \Delta x.
\label{eq:res}
\end{equation}

On the other hand, in the nonhydrostatic
Boussinesq system, when fronts occur, their (vertical to horizontal)
aspect ratio scale as $f/N_0$ \cite[e.g.,
see][]{snyder1993frontal}. When frontal collapse occurs, there
is an cascade of energy to grid scales, impeded only
by the small scale viscosity and diffusion (parameterized here by
hyperviscosity and hyperdiffusivity). Therefore, if fronts are to be
adequately resolved, vertical resolution will again have to
satisfy requirement (\ref{eq:res}). As pointed out in these
references, if the resolution requirement (\ref{eq:res}) is not met,
spurious dynamics including spurious gravity wave activity can occur and
incorrect numerical solutions are possible. All numerical
simulations presented here satisfy the above resolution
requirement. 

\subsection{Small scale dissipation}
If an isotropic Laplacian dissipation operator were to be used, as
in the Navier-Stokes equations, the prognostic governing
equations would be
\begin{gather}
{\partial \vv{u}_h \over \partial t} + \vv{u} \bcdot \bnabla \vv{u}_h
+  \frac{\vv{k} \times \vv{u}_h}{\Ro} = -\frac{\bnabla_{\!h} \phi}{\Ro}
+ \frac{1}{\RN} \left( \bnabla_h^2{\vv u_h} + \frac{1}{\AR^2}\frac{\partial^2 \vv
    u_h}{\partial z^2}\right)
\cr
\AR^2
\left({\partial w \over \partial t} +
  \vv{u} \bcdot \bnabla w \right) =  
-\frac{1}{ \Ro} \frac{\partial \phi}{\partial z} + \frac{b}{\Fr} 
+ \frac{\PR}{\RN} \left( \AR^2 \bnabla_h^2{w} + \frac{\partial^2 w}
  {\partial z^2}\right)
\cr
{\partial b \over \partial t} + \vv{u} \bcdot \bnabla b +  \frac{w}{\Fr}
=  
+ \frac{1}{\RN} \left( \bnabla_h^2{b} + \frac{1}{\AR^2}\frac{\partial^2 b}{\partial
    z^2}\right)
\label{eq:bouss_dssp}
\end{gather}
where $\RN$ is the Reynolds number and $\PR$ is the Prandtl ratio, the
ratio of buoyancy diffusivity to viscosity.  In each of the three
equations above, the ratio of vertical dissipation to horizontal
dissipation is $\frac{1}{\AR^2}$.  Thus, with the small aspect ratio
domains that we are interested in (atmospheres and oceans), vertical
dissipation is seen to dominate horizontal dissipation.  Furthermore,
the dynamical scales of interest in this study are in the range of
hundreds of meters to tens of kilometers, many orders of magnitude
larger than the Kolmogorov dissipation scales at which an isotropic
Laplacian dissipation would be appropriate. For these reasons, the
computations are more in the nature of Large Eddy Simulations rather
than Direct Numerical Simulation of the Navier-Stokes
equations. Furthermore, in atmosphere and ocean settings, at the
scales of interest, diapycnal mixing coefficients are orders of
magnitude smaller than isopycnal mixing coefficients.  Therefore, as
is common practice \citep[e.g.,][]{molemaker2010balanced} we choose a
simple hyperviscous but anisotropic dissipation operator that
localizes dissipation to the smallest of the resolved scales and set
the (eddy) Prandtl ratio to unity. The dissipation operator is
therefore of the form
$$\nu \left(\bnabla_h^2 + \frac{\nu_v}{\nu} \frac{\partial^2 }{\partial
    z^2}\right)^n.$$ 

In order to confine the effects of dissipation to the smallest of the
resolved scales and to extend the inertial range, an eighth order
hyper-diffusion operator was used for the dissipation of both momentum
and buoyancy.  For the main sequence of simulations considered in the
article $\nu$ in the above equation is determined so that the
hyperviscous-Reynolds number based on the grid scale is no
more than four. For the reference case with $\Ro=0.03$, the simulation
was repeated with various variations to the small-scale dissipation operator to study
dependence of results on them. For example,
Fig.~\ref{fg:dssp_verify} shows dissipation spectra and spectral flux of
kinetic energy in two such simulations with increased viscous and
diffusive coefficients (grid Reynolds number less than 1.6 and
0.4). There is no discernible trend in the dissipation spectra for the
three cases other than the location of the peak itself and the
differences are small. Further, over
the range of wavenumbers where there is a forward cascade of energy in
the main panel of Fig.~\ref{fg:dssp_verify}, while the least dissipative
case (the reference case) shows a larger flux at both the lower and
upper ends of this range, the trend is reversed over the intermediate
range of scales. Further, in otherwise identical experiments, we have
experimented with the scaling used in \cite{chasnov1994similarity} to
determine the value of $\nu$, and found the two methods to lead to
remarkably similar temporal profiles of dissipation rate. These
results suggest the robustness of the results we present to the
parameterization of small-scale dissipation.

Finally, we note that inevitable artifacts of finite resolution lead
to a certain fraction of energy cascading forward \citep[e.g. see
footnote on page 264 in][for an estimate of
this fraction]{nadiga2010alternating}. This is evident in
simulations of (interior) quasi-geostrophic turbulence (or
two-dimensional turbulence), which dynamics is by definition
balanced. We will account for this effect approximately by considering
dissipation and dynamics of twin simulations using the quasi-geostrophic
approximation and comparing results. 

\begin{figure}
\centering
\includegraphics[trim=0 0 0 0, clip, width=0.49\textwidth]{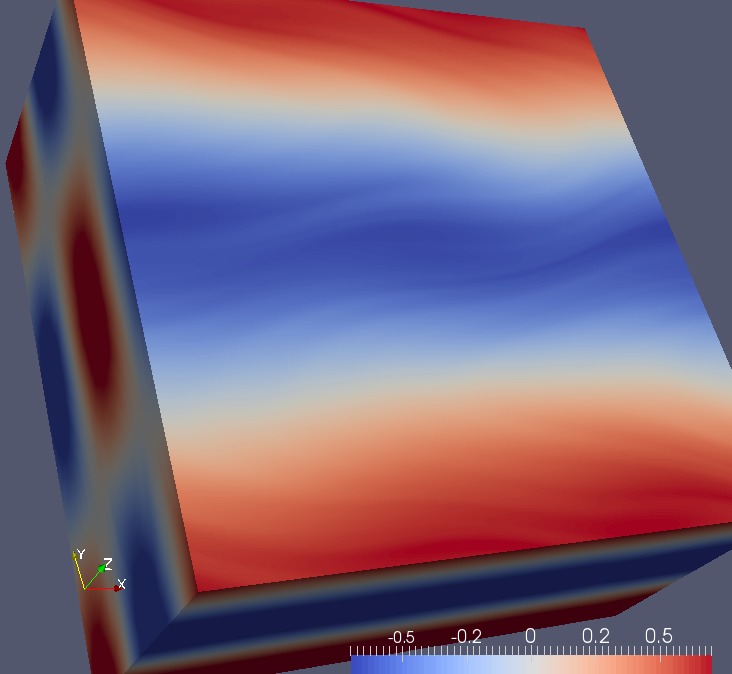}
\includegraphics[trim=0 0 0 0, clip,
width=0.49\textwidth]{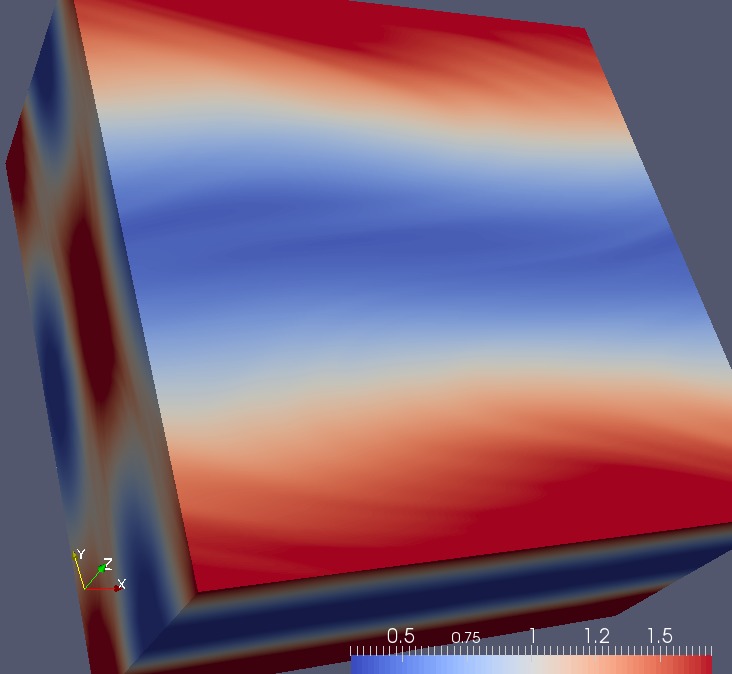}
\caption{Growth of the unstable baroclinic wave at time 1.2 for the QG
  system in the left panel and for the Boussinesq system in the right
  panel.  Potential vorticity is plotted.}
\label{fg:pv-40}
\end{figure}

\begin{figure}
\centering
\includegraphics[width=\textwidth]{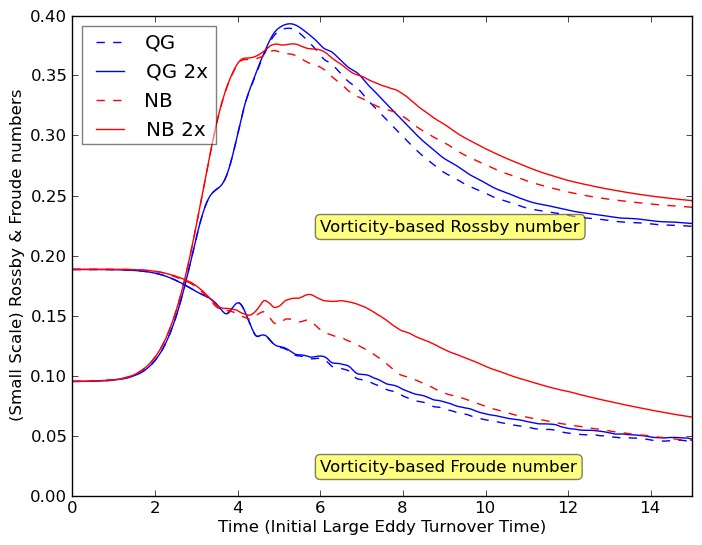}
\caption{Evolution of the rms of the vorticity-based Rossby and Froude
numbers ($(\zeta/f)_{(rms)}; (\omega_h/N_0)_{(rms)}$) under the QG and
Boussinesq approximations. Initial evolution 
is seen to be similar under the two approximations and at different
resolutions. Evolution in the later part of the instability phase is
seen to be different in the two systems with suggestions of an
instability other than the initial geostrophic hydrostatic baroclinic
instability in the Boussinesq system.}
\label{fg:rms_Ro_Fr}
\end{figure}

\section{Nonlinear Flow Evolution}
Our interest is in the parameter regime where the velocity-based $\Ro$
and the (vertical) velocity-based $\Fr$, both based on the domain
scales are small and the domain itself is such that the aspect-ratio
$\AR$ is small. Further, we will hold the ratio $\Fr/\Ro$, which is
the ratio of the horizontal size of the domain to the first internal
Rossby deformation radius, fixed at two. (The most unstable
  wavelength in this setup is shorter than the corresponding
  wavelength in the Eady problem by a factor of about two.)  Based on
these considerations, we choose a reference set $\AR = 0.05$, $\Ro =
0.03$, and $\Fr = 0.06$. We are in the process of conducting linear
stability analysis of the particular flow configuration used in this
study and will report on it in the near future.

Initial conditions for these values of the parameters are shown in
Fig.~\ref{fg:IC}. The color scale for potential vorticity in the QG
and Boussinesq systems are chosen to be symmetric about the ambient
value. Nevertheless, visual differences in the potential vorticity
field between the two system are evident if the lower panels of
Fig.~\ref{fg:IC}. For this reason, we caution the reader that spatial plots
of fields should be used only as an aide in visualizing flow and 
qualitative differences in evolution between the two systems.

We now consider the evolution of the above initial conditions under
both quasi-geostrophic dynamics and the non-hydrostatic Boussinesq
dynamics.  Further, in order to establish the role of resolution, we
consider evolution under the QG approximations at two different
resolutions and that under the Boussinesq equations at three different
resolutions. The two common resolutions correspond to spanning the
(nondimensional) unit length in the horizontal by 480 (1x) and 960 (2x)
points. The third resolution considered for the Boussinesq system is
intermediate between the above two resolutions: 720 (1.5x). At each of
these horizontal resolutions, the vertical resolution is chosen to
satisfy requirement (\ref{eq:res}.

With the above choice of initial conditions that consisted of a
unidirectional base flow and perturbations, each of which was
individually balanced, the unidirectional base flow was readily
destabilized and the process of geostrophic hydrostatic baroclinic
instability unfolded. Figure~\ref{fg:pv-40} shows the growth of the
unstable baroclinic wave at time 1.2 for the QG system in the left
panel and for the Boussinesq system in the right panel; in these
figures potential vorticity is plotted.  We note here that time is
non-dimensionalized by the initial eddy turnover time and that the
characteristic eddy turnover at later times is less than the chosen
initial eddy turnover time.  Besides similarities in the gross flow
structure, what is clear in this figure is that there are significant
differences in the growing perturbations even at this early
time. However, a characterization of these differences will have to
await the completion of linear stability analysis presently underway.

\subsection{Evolution of vorticity-based Rossby and Froude numbers and
other relevant limits}
The velocity-based Rossby and Froude numbers of the initial conditions
serve, at best, as a rough guide and could in some cases be misleading
(as, e.g., in a setup where the domain scale is much greater
than the deformation radius). For this reason, we compute the
vorticity-based Rossby and Froude numbers; not only do they serve as
dynamically relevant counterparts of the velocity-based ones, but they
also emphasize the smaller scales. They are defined respectively as
\begin{gather}
\Ro_\omega = \frac{\zeta}{f},
\label{eq:ro_zeta}
\end{gather}
where $\zeta$ is the vertical component of relative vorticity and 
\begin{gather}
\Fr_\omega = \frac{|\omega_h|}{N_0},
\label{eq:fr_omega}
\end{gather}
where $|\omega_h|$ is the magnitude of the horizontal components of
relative vorticity.  Figure~\ref{fg:rms_Ro_Fr} shows the evolution of
rms values of vorticity-based Rossby and Froude
numbers. $\Ro_\omega^{(rms)}$ is seen to remain at its initial value
over the first couple initial-eddy-turnover times, after which, it
begins to increase as the unstable baroclinic wave begins to wind
up. In both systems, $\Ro_\omega^{(rms)}$ reaches its peak value of
about 0.4 at about time 5 after which it begins to fall. While there
is overall similarity in the evolution of $\Ro_\omega^{(rms)}$ in the
two systems, it is to be noted that a) the peak value of
$\Ro_\omega^{(rms)}$ is higher in the QG system than in the Boussinesq
simulations whereas at late times $\Ro_\omega^{(rms)}$ is lower in the
QG system, and b) the ascent and descent around the peak is temporally
more extended in the Boussinesq system. As to whether the latter
behavior is due to the possibility of other instabilities than the
hydrostatic geostrophic baroclinic instability in the Boussinesq
system unlike in the QG system is not clear.  

$\Fr_\omega^{(rms)}$ is seen to display a decreasing trend through the
instability phase although the behavior is not monotonic.  This
behavior is consistent with the conversion of the vertical shear of
horizontal velocity (main component of horizontal vorticity) into
horizontal shear of horizontal velocity associated with baroclinic
instability. Further, $\Fr_\omega^{(rms)}$ is seen to higher in the
Boussinesq system most of the time. Two possible scenarios that could
result in such larger values are a) If the horizontal vorticity is
balanced, then it could be that the conversion of
initial potential energy into kinetic energy is smaller in the Boussinesq
system, or b) there is a significant amount of unbalanced horizontal vorticity.

Finally, we note that while there is overall agreement in the behavior
of $\Ro_\omega^{(rms)}$ and $\Fr_\omega^{(rms)}$ between the
simulations at the two resolutions, there is a greater dependence on
resolution in the Boussinesq system.

\begin{figure}
\centering
\includegraphics[width=\textwidth]{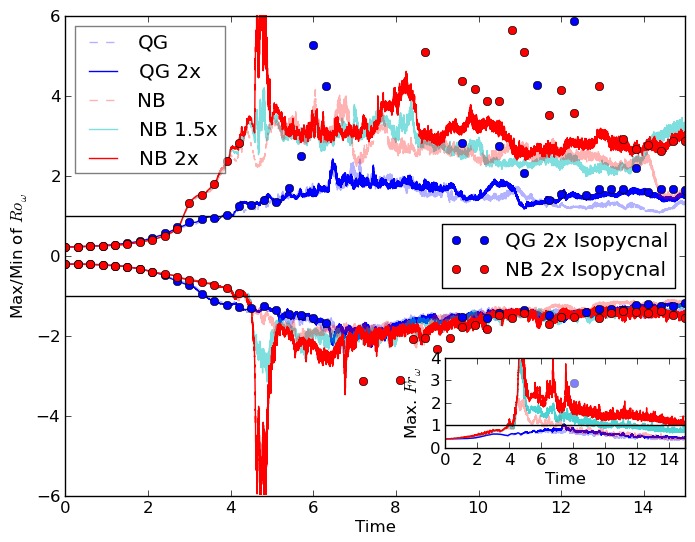}
\caption{Evolution of the minimum and maximum of the vorticity-based
  Rossby number ($\zeta/f$) suggests spatially and temporally
  intermittent events that lead to $O(1)$ $\Ro_\omega$ locally. While
  lines show $\Ro_\omega$ values computed using horizontal derivatives
  at fixed $z$ locations, filled circles show values computed using
  horizontal derivatives on isopycnal surfaces.$\Ro_\omega$ falling
  below -1 in the QG system is equivalent to violating second
  balanced-integrability requirement as reviewed in Molemaker et al.,
  2005. Asymmetry between cyclonic and anticyclonic events is evident
  in the Boussinesq system. Simulations at differing resolutions
  suggest an increase in intermittent events as resolution increases
  in the Boussinesq system.  In the inset is shown the maximum of the
  absolute value of the vorticity based Froude number.}
\label{fg:minmax_rv}
\end{figure}

\begin{figure}
\centering
\includegraphics[width=\textwidth]{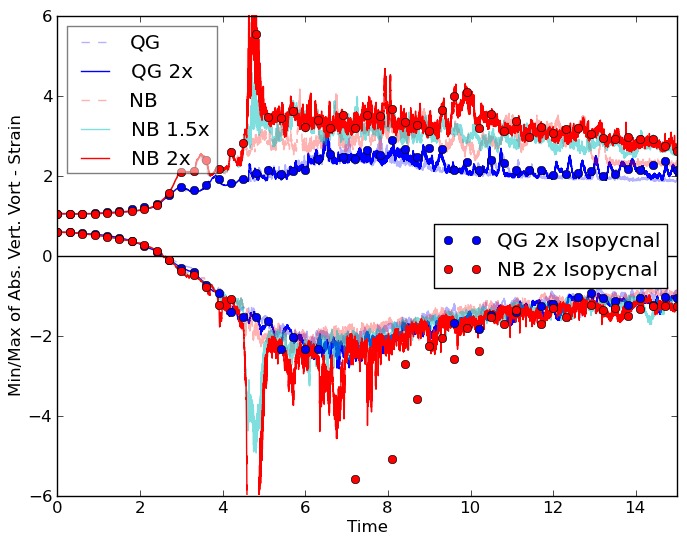}
\caption{Evolution of peak values of the difference between absolute
  vorticity and strain. Change in sign of this quantity in the QG
  approximation, if only locally, suggests in-applicability of that
  approximation in those regions. There is general agreement in the
  evolution of this quantity in the QG and Boussinesq approximations.}
\label{fg:ams}
\end{figure}

\begin{figure}
\centering
\includegraphics[width=\textwidth]{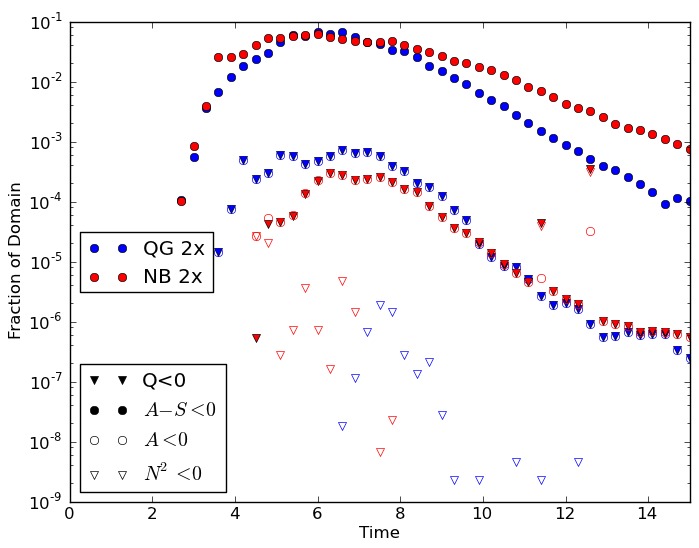}
\caption{Evolution of the fraction of domain where certain dynamical
  limits are crossed in the QG and Boussinesq systems. The limits are
  indicated in the lower legend and correspond to (a) negative
  potential vorticity, (b) $A-S < 0$ where $A$ is absolute vertical
  vorticity and $S$ is the magnitude of strain as defined in
  (\ref{eq:mcw2}), (c) $A<0$, and (d) $N^2<0$.}
\label{fg:fraction}
\end{figure}

\citealp{mcwilliams1998breakdown} consider the solvability limits
of a version of balance equations formulated in isentropic coordinates
(see references therein) and conjecture that those limits coincide
generally with the boundary between distinctive nonlinear dynamical
behaviors associated with large-scale and small-scale regimes.  The reader
is referred to the above article for details; the limits are also
reviewed in Molemaker et al., 2005. In brief, the above limits are
related to changes in sign of a) stratification (gravitational
instability), b) absolute vorticity A (centrifugal or inertial
instability):
\begin{gather}
A<0, \quad A = f + \zeta,
\label{eq:mcw1}
\end{gather}
and c) the difference between absolute vorticity and the horizontal
strain rate S (anticyclonic ageostrophic baroclinic instability):
\begin{gather}
 A-S<0, \quad S = \sqrt{(u_x - v_y)^2 + (v_x + u_y)^2}.
\label{eq:mcw2}
\end{gather}
They further suggest that these limits are likely, approximately,
indicative of the transition from the larger-scale regime of inverse
energy cascades by anisotropic flows to the smaller-scale regime of
forward energy cascade to dissipation by more isotropic flows
and intermittently breaking inertia-gravity waves. We also note here
that there is some work to suggest that the last two of the above
kinds of instability may be dynamically related \cite[e.g.,
see][]{bouchut2011inertial}, but defer an examination of this issue to
the future. We note that all initial conditions in this study
are such that all the conditions for integrability are satisfied.

The initial conditions are such that the occurrences of unstable
stratification is rare (see Fig.~\ref{fg:fraction}) and 
in the rare occasion it occurs it is likely caused by a shear
instability. We then consider the violation of the other two
integrability conditions (\ref{eq:mcw1}) and (\ref{eq:mcw2}).  
Figure~\ref{fg:minmax_rv} displays the
evolution of the minimum and maximum of the vorticity-based Rossby
number over the same period; for reference, the minimum and maximum
for evolution under the Boussinesq approximation is given as well. The
dashed and continuous curves correspond to values of $\Ro_\omega$ where
the horizontal derivatives are computed at fixed $z$. On the other
hand, filled circles are corresponding values of $\Ro_\omega$ computed
on isopycnals. That is if $u_X$ is the horizontal derivative on an
isopycnal, then
$$u_X = \left.\frac{\partial u}{\partial x}\right|_{B=cst.} 
= u_x - u_z \, \frac{\partial B/\partial x}{\partial B/\partial z},$$
where $B$ is total buoyancy (or density) that is evolving.
The isopycnal results are shown every 0.3 time units. Up to times when
peak values of $\Ro_\omega$ first cross unit magnitude, the values of
$\Ro_\omega$ computed on the horizontal and isopycnal match well. This
is largely the case at late times (t$>$12) as well; large differences
are confined to intermediate times.

In Fig.~\ref{fg:rms_Ro_Fr}, the domain
averaged (rms) values of $\Ro_\omega$ are seen to remain less than
0.4. However, Fig.~\ref{fg:minmax_rv} shows an extended period of time
when the peak values of
$\Ro_\omega$ are greater than unity in the QG approximation (likewise
with the Boussinesq system). This is suggestive of spatially and
temporally intermittent events that lead to $O(1)$ $\Ro_\omega$
locally. The occurrence of $O(1)$ $\Ro_\omega$ in the quasi-geostrophic
system in localized regions is indicative of the in-applicability of
the approximation in those regions and in particular, the second
condition for the loss of integrability,  equivalent to the
minimum of $\Ro_\omega$ falling below unity, is seen to be met
in the QG simulations, first at a time close to 3.3.

Further, we note that
a) In contrast to the behavior of the rms, peak values of
  $\Ro_\omega$ are higher in the Boussinesq system suggesting 
  greater intermittency in the Boussinesq system compared to the QG system.
b) Increased resolution leads to higher levels of intermittency in
  the Boussinesq system.
c) A greater degree of asymmetry is seen between cyclones and
  anticyclones in the Boussinesq system---peak cyclonic values are
  stronger than peak anticylonic values.
d) The QG approximation reproduces the minimum value of $\Ro_\omega$
(peak anti-cyclonic values) found in the Boussinesq system remarkably
well other than in (likely) frontal regions. It fares much worse in
the cyclonic region.

  In order to better diagnose dynamics and for future reference,
  Fig.~\ref{fg:minmax_rv} shows in the inset plot the time evolution
  of the peak value of the vorticity based Froude number
  $\Fr_\omega$. In this figure, the time of occurrence of maxima may be
  seen to be approximately the same as the time of occurrence of
  extrema in $\Ro_\omega$. While the peak value in the QG system is
  close to one, much higher values are attained in the Boussinesq
  system. Comments about increased intermittency in the Boussinesq
  system made in the context of $\Ro_\omega$ hold for the $\Fr_\omega$
  diagnostic as well.

We also monitor $A-S$ as relates to the third of the
balanced-integrability conditions above. Figure~\ref{fg:ams} shows the
evolution of the peak values of $A-S$. Again, while lines show
values obtained with horizontal derivatives evaluated at constant $z$,
circles show values obtained using horizontal derivatives on isopycnal
surfaces. For this diagnostic, differences between the two values are
large only in anticyclonic regions at intermediate times. 

The change in sign of $A-S$ as the baroclinic instability unfolds in
the QG system further confirms the violation of the integrability
condition for the QG system in localized regions seen in the previous
diagnostics. Indeed, the requirement that the change in sign of this
diagnostic should occur earlier than the violation of the second
solvability condition is confirmed: $A-S$ changes sign at a time close
to 2.5 where as $A$ changes sign at around 3.3. Further, it is
possible that the increase in the maximum of $\Ro_\omega$ in the
Boussinesq system at time 2.5 is related to the change in sign of
$A-S$. These diagnostics are then consistent with the possibility that
the dynamical conditions are conducive for the occurrence of the
anticyclonic ageostrophic baroclinic instability of
\citealp{mcwilliams1998breakdown} (as distinct from gravitational and
centrifugal instabilities) in the context of a set of governing
equations that do not make the geostrophic balance approximation.

Next, Fig.~\ref{fg:fraction} shows fractions of the domain where the
three solvability limits are violated for the QG system as a function
of time. Also shown is the fraction of domain where quasi-geostrophic
potential vorticity falls below -1 (indicated as $Q<0$ where $Q_{qg} =
1+q$ and $q$ is given by (\ref{eq:qg})). Here it is seen that the
violation of the $A-S$ criterion is most extensive and reaching a
value of about 8\% of the domain at around time 6.5. As mentioned
earlier, that this condition is violated most extensively is expected
and related to the fact that this condition occurs at a smaller value
of the local Rossby number as compared to, say, the absolute
vorticity criterion. The latter $A<0$ is next most often violated with
violations of static stability being very infrequent.  Further, a very
high degree of correlation is seen between the fraction of the domain
where potential vorticity is negative and where absolute vertical vorticity is
negative suggesting that the flow may be susceptible to centrifugal and
inertial instabilities. 

Given the conjecture of \citealp{mcwilliams1998breakdown} of the
relevance of these limits to the transition from the larger-scale
regime of inverse energy cascade to the smaller-scale regime of
forward energy cascade, fractions of the domain where these conditions
are met in the Boussinesq system are shown in Fig.~\ref{fg:fraction}
as well. This will be useful later in interpreting dynamics in the
Boussinesq system. We note, however, that there is generally a good
correspondence in this diagnostic between the two systems. In
particular, the strong correlation seen between the fraction of the domain
where potential vorticity is negative and where absolute vertical vorticity is
negative in the QG system holds in the Boussinesq system as well.

\begin{figure}
\centering
\includegraphics[trim=0 0 0 0, clip,
width=0.49\textwidth]{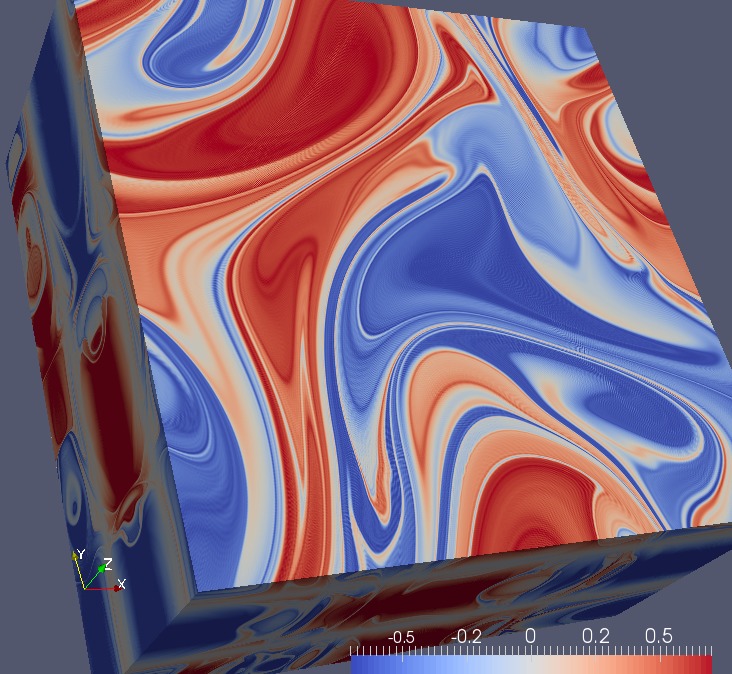} 
\includegraphics[trim=0 0 0 0, clip,
width=0.49\textwidth]{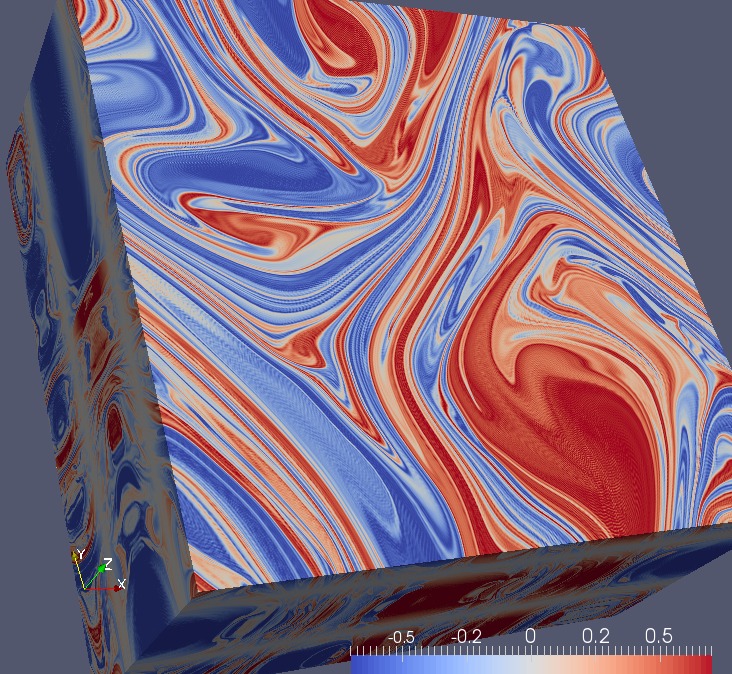}
\includegraphics[trim=0 0 0 0, clip,
width=0.49\textwidth]{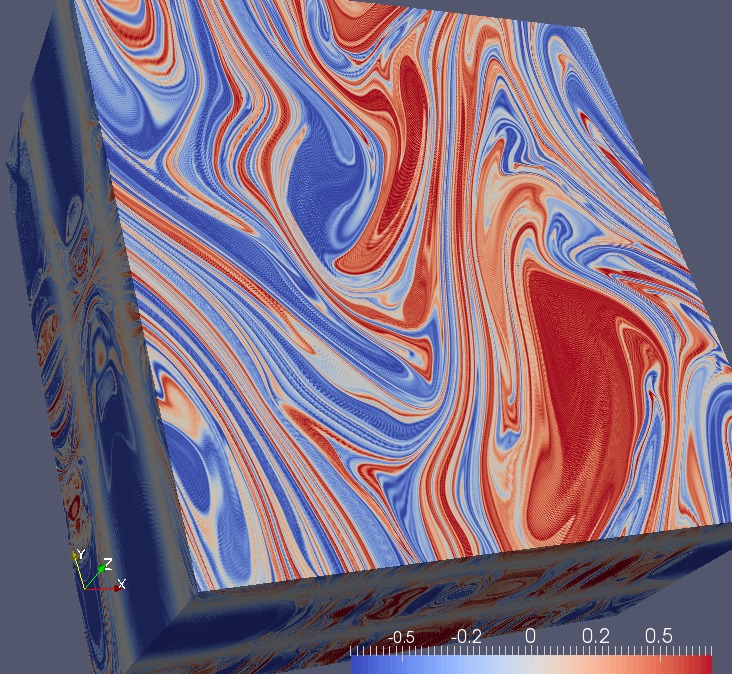}
\includegraphics[trim=0 0 0 0, clip,
width=0.49\textwidth]{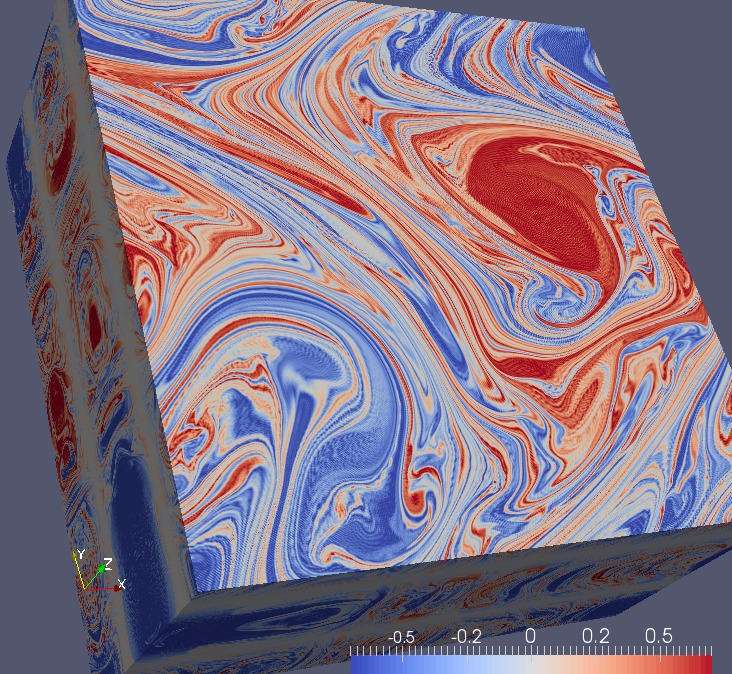}
\caption{Snapshots of perturbation potential vorticity in the QG
  system. The times of the snapshots are close to the time of:
  top-left: the local peak in $\Ro_\omega$ around 4.8; top-right:
  global minimum of $A-S$ around 6.1; bottom-left: global minimum of
  $\Ro_\omega$ and global maximum of $\Fr_\omega$ near 6.4;
  bottom-right: global maximum of $\Ro_\omega$, around 7.0.}
\label{fg:qgpv}
\end{figure}

\subsection{Evolution in the QG System}
Given the small Rossby and Froude numbers of the initial condition in
the thin aspect ratio domain we consider, the general aspects of the
flow evolution under the QG and Boussinesq systems are similar and
consistent with those expected from geostrophic turbulence
theory. Initial evolution of the smooth initial condition consists of
a linear phase when the most unstable baroclinic perturbation grows.
Subsequent (nonlinear) evolution of the instability leads to a roll-up
into eddies. The break-up of the flow and subsequent nonlinear
evolution leads to an inverse cascade of energy that results in a
build up of energy in the barotropic mode. Energetically the large-scale aspect
of flow evolution comprises a conversion of available potential energy
of the initial condition to eddy kinetic energy followed by an inverse
cascade of kinetic energy. 

In order to convey a qualitative picture of the flow evolution, Figure~\ref{fg:qgpv}
shows the distribution of potential vorticity in the QG system at
a few selected times. Looking back at Figs.~\ref{fg:minmax_rv} and
\ref{fg:ams}, we choose the following four times: a) time 4.8 which is
close to the time of a local peak of maximum $\Ro_\omega$, b) time 6.1 which is
close to the time of global minimum of $A-S$, c) time 6.4 which is
close to the time of both the global minimum of $\Ro_\omega$ and
global maximum of $\Fr_\omega$, and d) time 7.0 which is close to the 
global maximum of $\Ro_\omega$. We note that the global maximum of the
rms of $\Ro_\omega$ occurs at time 5.2 which is close to the time of
the top-left snapshot.

\begin{figure}
\centering
\includegraphics[width=\textwidth]{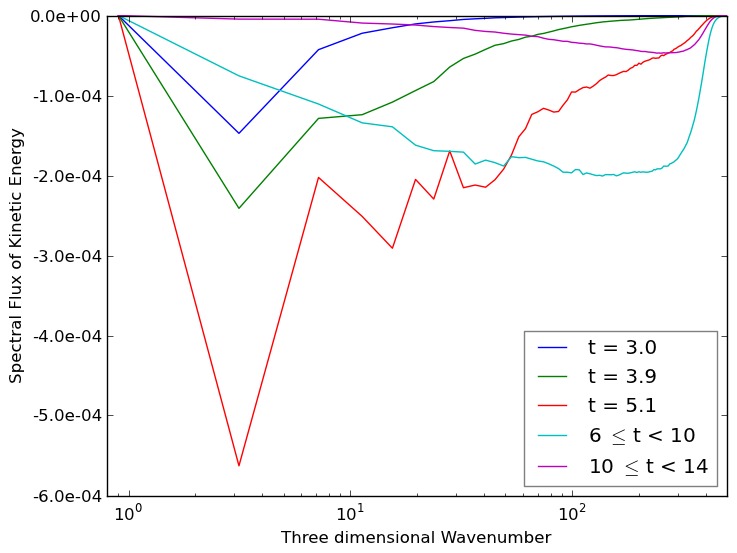}
\caption{A plot of the nonlinear spectral flux of kinetic energy as a
  function of three-dimensional wavenumber in the QG 2x case. Three
  instantaneous plots over the early period and two time-averaged
  plots over intermediate and late times are show. A progressive
  destabilization of  small scales is seen. All instabilities,
  however, lead to an inverse cascade of kinetic energy.}
\label{fg:qgflux}
\end{figure}

Over the range of times considered, a progressive destabilization of
smaller scales of the flow with time is evident from these snapshots
and they may be thought of as instabilities associated with various
baroclinic modes of the system. We note here that given the
spatially-variable nature of the vertical shear of zonal velocity and
stratification in the base flow, the structure and scale of these
baroclinic modes are not immediately apparent; linear stability
analysis that is ongoing is likely to shed more light on these
aspects.  The progressive destabilization of small scales can also be
seen in the plot of spectral flux of energy at different times in
Fig.~\ref{fg:qgflux}. However, and as expected, all of these
instabilities only lead to an inverse cascade.  Additionally, we have
previously seen that the QG approximation tends to be inapplicable in
at least some isolated regions of the flow domain beyond times as
early as 2.5. While isolating such regions and examining dynamics in
their vicinity could lead to further insight, we have not attempted
this yet.

\subsection{Front Formation in the Boussinesq System}
\label{sec:front}
\begin{figure}
\centering
\includegraphics[width=\textwidth]{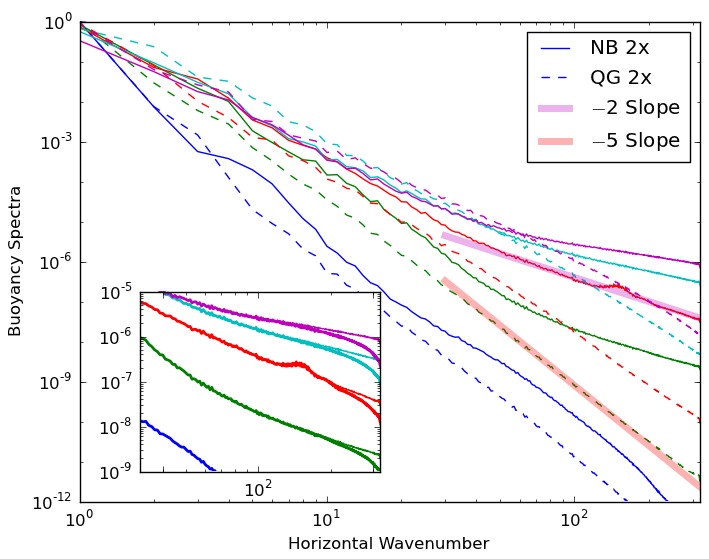}
\caption{Buoyancy spectrum is shown as a function of horizontal
  wavenumber for the QG and Boussinesq systems at times 2.4 (blue),
  3.6 (green), 4.5 (red), 6.0 (cyan), and 8.0(magenta). Spectral
  slopes past time 2.4 are steeper in the Boussinesq system than in
  the QG system suggesting occurrence of fronts in the Boussinesq
  system. Slopes close to $-2$ are seen at some of the intermediate
  times such as seen here at time 4.5. Close spacing of spectra at
  times 6 and 8 suggest a slowdown in frontogenesis past a time of
  about 6. The inset plot shows that there are minor changes to the
  spectra at wavenumbers smaller than about 200 when hyperviscosity
  and hyperdiffusivity coefficients in the horizontal are increased by
  a factor of 10.}
\label{fg:ape-spec}
\end{figure}

As mentioned in the previous subsection, there is general similarity
in the initial evolution of the Boussinesq system to that in the QG
system. Past time 2.5, however, evolution of the two systems differ
significantly and this is evident in previously presented diagnostics
such as those in Figs.~\ref{fg:rms_Ro_Fr}, \ref{fg:minmax_rv}, and
\ref{fg:ams}.  Indeed a striking difference in the evolution of the
Boussinesq system as compared to that of the QG system is the
spatially-localized event around time 4.8 (starting at about time 4.5
and ending at about time 5.1) seen in Fig.~\ref{fg:minmax_rv} when peak
vorticity-based Rossby and Froude numbers briefly reach large
values. This event is not seen in the domain-averaged measures such as
in Fig.~\ref{fg:rms_Ro_Fr}. It is possible that this event is related
to the formation of steep fronts by mesoscale shearing and
straining. To further explore this possibility, Fig.~\ref{fg:ape-spec}
shows buoyancy spectra at times 2.4, 3.6, 4.5, and 6.0 for the
quasi-geostrophic system and the Boussinesq system in the main
panel. The fall-off of the spectra in the quasi-geostrophic system is
seen to be close to the expected $-5$ scaling. The spectra of the
Boussinesq system (at $t>$2.4) are seen to be steeper. Indeed, their
slope is seen to be close to $-2$, as would be expected with the
formation of fronts, and somewhat shallower than the -8/3 slope
obtained by \cite{andrews1978energy} using semigeostrophic theory (For
why the spectral slope tends to more often be closer to $-2$ than
$-8/3$, see \cite{boyd1992energy}.)

Next, compare the difference in the spectra at small scales in the two
systems between times 2.4 and 3.6. The accumulation of energy at small
scales is seen to be significantly faster in the Boussinesq system
than in the QG system. This is most likely related to the influence of
ageostrophic circulation of the time scale for frontogenesis
\citep{holton2004dynamic}. This influence in the Boussinesq system
is like in the semigeostrophic system that is typically used for
analyzing frontogenesis \cite[e.g.,
see][]{holton2004dynamic}. This influence leads to a positive
feedback in the growth of horizontal density gradients because an
increase in the horizontal density gradient leads to an increase in
the strength of the secondary ageostrophic circulation which in turn
accelerates the growth of horizontal density gradient. However, in the
QG system, with only the geostrophic circulation driving the increase
in horizontal density gradient, this feedback is absent and the
resulting rate of growth in horizontal density gradient is smaller.
Further, since the formation of fronts is likely an important process
in the chain of events leading to dissipation of balanced mesoscale
energy, the dependence of the buoyancy spectra shown in the main panel
for the Boussinesq system to a change in the coefficient of
small-scale dissipation parameterization is shown in the inset. In
this inset, the thicker lines correspond to a case where the
coefficient of horizontal hyperviscosity and hyperdiffusivity have
been increased by a factor of 10. Changes to spectra at wavenumbers
smaller than about 200 are seen to be minimal. 

\begin{figure}
\centering
\includegraphics[trim=0 0 0 0, clip,
width=0.49\textwidth]{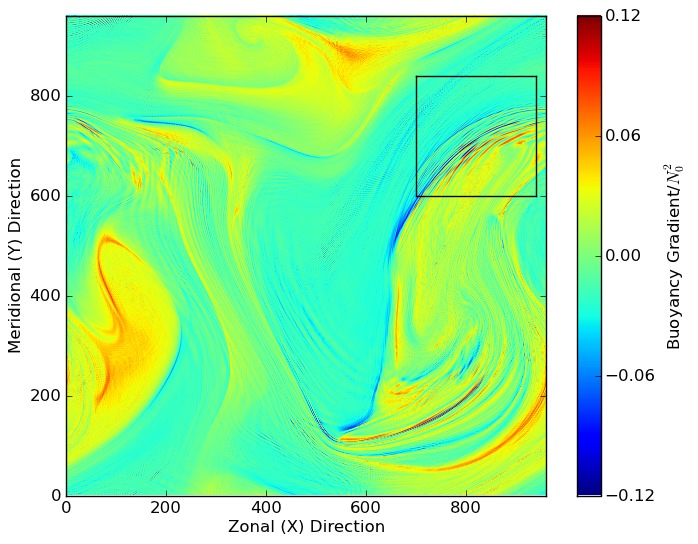}
\includegraphics[trim=0 0 0 0, clip,
width=0.49\textwidth]{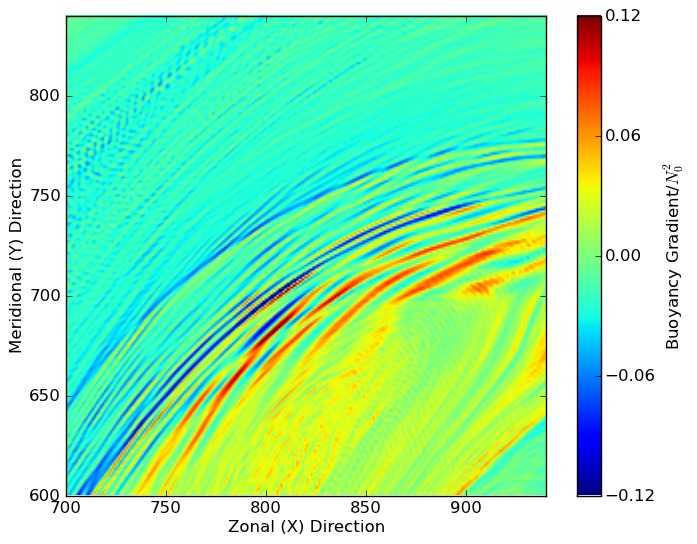}
\caption{Magnitude of the horizontal gradient of buoyancy on an $x-y$
  plane at time 4.5. The full $x-y$ domain is shown on the left. A
  zoom of the region marked by the square is shown on the right. The
  $x-y$ plane corresponds to the top surface in Figs.~\ref{fg:IC},
  \ref{fg:pv-40}, and \ref{fg:qgpv}}
\label{fg:front}
\end{figure}

Figure~\ref{fg:front} shows the horizontal gradient of buoyancy at
time 4.5 when a local peak in the buoyancy spectrum is seen in
Fig.~\ref{fg:ape-spec} at a horizontal wavenumber of about 150. The
$x-y$ plane corresponds to the top surface in Figs.~\ref{fg:IC},
\ref{fg:pv-40}, and \ref{fg:qgpv}. The left panel shows the full $x-y$
plane and the right panel shows a zoom of the region marked by the
rectangle in the left panel. Elongated frontal features with narrow
cross-frontal widths seen in this figure are likely regions of
enhanced unbalanced instabilities. We also note that regions of strong
horizontal density gradients did not preferrably coincide with regions
of either weak or strong stratification (e.g., see
Fig.~\ref{fg:pvminloc}). leading to an irreversible forward cascade of
energy. While further investigation of the secondary ageostrophic
circulation in the cross-frontal plane may be possible, for example by
using the Sawyer-Eliassen equation, the curved nature of the fronts
makes this task difficult.

\subsection{Inertial and Symmetric Instability}
We now consider the possibility of inertial and symmetric
instabilities occurring in the Boussinesq system. To this end, we
briefly consider the dynamics of potential vorticity following
\cite{thomas2013symmetric}. For this purpose, it is useful to consider
potential vorticity as a combination of two terms, the first of which
comprises terms involving the vertical component of absolute vorticity
(and/or vertical gradients of buoyancy) whereas the second comprises
terms involving the horizontal component of vorticity (and/or
horizontal gradients of buoyancy). For this purpose, (\ref{eq:ndpv})
may therefore be rewritten as
\be
Q = Q_{vert}  + Q_{bc} = \left(1 + \Ro\zeta \right) 
\left( 1 + \Fr\frac{\partial b}{\partial z}\right) 
+ \bomega_h\cdot\nabla_h b.
\label{eq:vert-bc}
\ee
Dimensionally this corresponds to 
$Q^{(d)} = Q^{(d)}_{vert} + Q^{(d)}_{bc} = \zeta^{(d)}_aN^2
+ \bomega_h^{(d)}\cdot\nabla_h^{(d)} b^{(d)}$, where $\zeta_a$ is the
vertical component of (dimensional) absolute vorticity.

When the sign of potential vorticity is the 
opposite of that of the Coriolis parameter, the flow is unstable to
overturning instabilities \citep{hoskins1974role}. Since these arguments
are more intuitive and traditional in the dimensional form, we will
use that and the corresponding nondimensional form will be indicated
in parentheses.  (Referring to (\ref{eq:vert-bc}), the above condition
occurs when $Q<0$.)

The main cause for low (negative) potential vorticity can be
associated with vertical vorticity (first term on the RHS),
stratification (second part of first term) or baroclinicity of the flow (last term).\\
a) When $N^2<0$ ($1 + \Fr{\partial b}/{\partial z} < 0)$, the
instability is the gravitational instability---an instability to
vertical displacement of a fluid particle.\\
b) When $N^2>0$, but $f\zeta_aN^2 < 0$ 
($1 + \Fr{\partial b}/{\partial z}>0
$, but $\left(1 + \Ro\zeta \right) \left( 1 + \Fr{\partial
    b}/{\partial z}\right) < 0$)
the resulting instabilities are termed inertial or centrifugal
instabilities---an instability to horizontal displacement of a fluid particle.\\
c) When $N^2>0$ and $f\zeta_aN^2 > 0$, but $fq < 0$, the resulting
instabilities are termed symmetric instabilities---an instability to a
slantwise displacement of a fluid particle, also thought of as an
isopycnal inertial instability.

Thus for symmetric instability to occur, $Q_{bc}<0$ and $|Q_{bc}| >
|Q_{vert}|$, indicating that symmetric instability 
is associated with strong baroclinicity.
To consider this case further, we note from
scale analysis that the horizontal component of vorticity is well
approximated by 
$$\bomega_h  \approx \left(-\frac{\partial v}{\partial z} \widehat{\vv{x}},\,
\frac{\partial u}{\partial z} \widehat{\vv{y}} \right).
$$
Consequently, the second term of (\ref{eq:vert-bc}) can be
approximated as 
$$
Q_{bc} \approx  
\frac{\partial u}{\partial z}  \frac{\partial b}{\partial y} - 
\frac{\partial v}{\partial z}  \frac{\partial b}{\partial x}.
$$
For a flow in thermal wind balance, this reduces to 
$$
Q^g_{bc} = -\frac{1}{f}|\bnabla_h b|^2,
$$
showing that baroclinicity results in lowering potential vorticity
\citep{thomas2013symmetric}. However, in an unbalanced flow,
baroclinicity could act to either decrease or increase potential
vorticity.

\begin{figure}
\centering
\includegraphics[width=\textwidth]{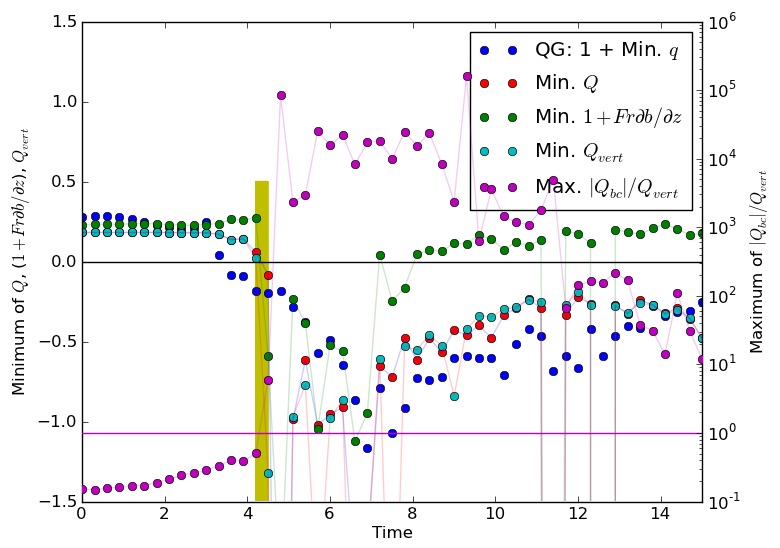}
\caption{Evolution of three (Ertel) potential vorticity related
  diagnostics are plotted as a function of time of the NB 2x
  case. These are the minimum value of potential vorticity in red,
  minimum value of the vertical component of potential vorticity as
  given by the first term on the RHS of (\ref{eq:vert-bc}) in cyan and
  the maximum value of the ratio of the absolute value of the
  baroclinic component of potential vorticity as given by the second
  term on the RHS of (\ref{eq:vert-bc}) and the vertical component
  (logarithmic scale on right) in magenta. Also shown are the the
  minimum of potential vorticity for the QG 2x case in blue and the
  minimum of stratification for case NB 2x in green. The yellow stick
  is placed to facilitate seeing the values of these terms just before
  (left edge of stick) and just after (right edge) the minimum value
  of potential vorticity becomes negative in case NB 2x. A 
  black horizontal line is drawn at the zero of the $y-$axis on the
  left and a horizontal line in magenta is drawn at at $y=1$ for the
  $y-$axis on the right.  Conditions for both inertial instability and
  symmetric instability are seen to be satisfied first at time 4.5
  and other subsequent times.}
\label{fg:pvmin}
\end{figure}

\begin{figure}
\centering
\includegraphics[width=\textwidth]{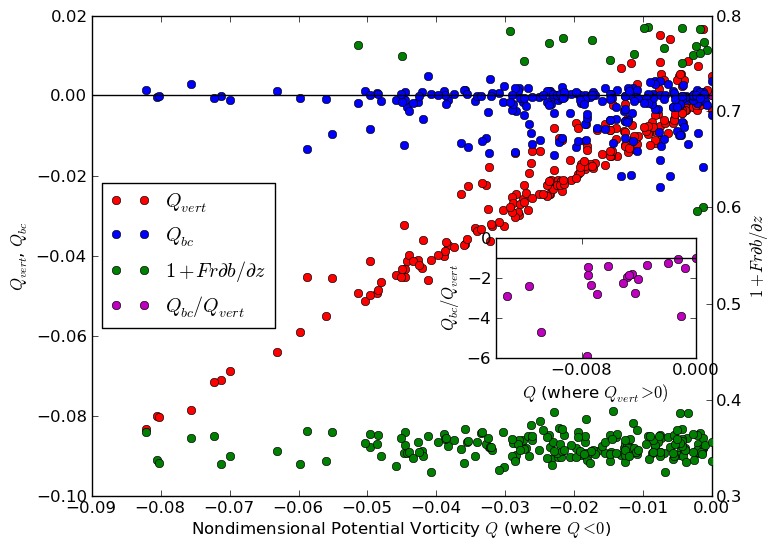}
\caption{Potential vorticity diagnostics at time 4.5 and at locations
  where $Q<0$. At each of
these points, values of $Q_{vert}$, $Q_{bc}$,
stratification, and the ratio $Q_{bc}/Q_{vert}$ are plotted against
potential vorticity; the last of these is
shown in an inset plot. At points where 
$Q_{vert}<0$ in the main panel conditions for
inertial instability are met. At points where $Q_{bc}/Q_{vert}<0$ in
the inset plot, conditions for symmetric instability are satisfied.}
\label{fg:pvmin150}
\end{figure}

We consider related diagnostics in Fig.~\ref{fg:pvmin}. In this
figure, the minimum of potential vorticity, minimum of stratification,
minimum of the $Q_{vert}$, and the maximum of
$|Q_{bc}|/Q_{vert}$ are shown as a function of time. For the first of
these three diagnostics, the minimum over the full domain is
considered. However, for the last of the diagnostic, the maximum is
considered over the part of the domain where $Q_{bc} < 0$, for the
reason that $Q_{bc}$ can be positive in an unbalanced flow. The
$y-$axis for the first of these diagnostics is on the left where as
that for the last is on the right side and a logarithmic scale is used.
A reference line at $y=0$ for the first three of the
diagnostics and a reference line at $y=1$ for the last diagnostic are
shown. A yellow stick is shown to highlight the first time potential
vorticity becomes negative in the flow (at $t\approx 4.5$ at the right
end of the stick) and the time preceding it (at $t\approx 4.2$ at the
left end of the stick).

At initial times minimum of $Q$ is coincident with the minimum of
$Q_{vert}$, the flow is stably stratified everywhere, and the maximum
of $|Q_{bc}|/Q_{vert}$ where $Q_{bc}<0$ is less than unity. 
Consider the state of the system at time 4.2 on the left side of yellow stick.
At this time, when for the first time, minima of $Q$ and $Q_{vert}$ are not
coincident, two conditions seem imminent: a) $Q_{vert}$ is tending to
go negative, and b) the magnitude of $Q_{bc}$ is tending to exceed the
value of $Q_{vert}$ in a region where $Q_{bc}<0$ and
$Q_{vert}>0$. These are exactly the conditions for a)
inertial/centrifugal instability and b) symmetric instability
respectively. 

Indeed at time 4.5 on the right side of the yellow stick,
potential vorticity has become negative for the first time and both of
the above conditions are seen to be satisfied (in separate regions),
suggesting that the flow is susceptible to both inertial/centrifugal
instability and symmetric instability. To verify this further, we
consider points in the domain at time 4.5 where $Q<0$. At each of
these points, we plot the values of $Q_{vert}$, $Q_{bc}$,
stratification, and the ratio $Q_{bc}/Q_{vert}$; the last of these is
shown in an inset plot. The nearly-diagonal distribution of
$Q_{vert}$ vs. $Q$ implies that $Q$ becoming negative is largely due
to $Q_{vert}$ becoming negative. That is, conditions for  inertial/centrifugal
instability are satisfied. 

Nevertheless, in regions where potential vorticity takes on small
negative values, it is seen that there are points with
$Q_{vert}>0$. Further, at these points, $Q_{bc}<0$ and the magnitude
of $Q_{bc}$ exceeds that of $Q_{vert}$ leading to conditions that
satisfy the necessary condition for symmetric instability. Further, an
examination of these diagnostics at subsequent times in conjunction
with those in Fig.~\ref{fg:pvmin} show that the conditions for these
instabilities are met over an extended period of time. 

The diagnostics in the previous section and this one then suggest the
following picture: Following the formation of mesoscale
eddies by hydrostatic geostrophic baroclinic instability, mesoscale
shear and strain drive the formation of fronts. Steep fronts first
form at a time of about 4.5. Potential vorticity
becomes negative for the first time at about the same time (hand in
hand). Thus, concurrent with the formation of
fronts, necessary conditions for inertial and symmetric instabilities
are also satisfied for the first time. These conditions continue
to be satisfied over most of the remainder of the time we consider
evolution over, whereas frontogenesis slows down beyond a time of
about 6. 

\begin{figure}
\centering
\includegraphics[width=.23\textwidth]{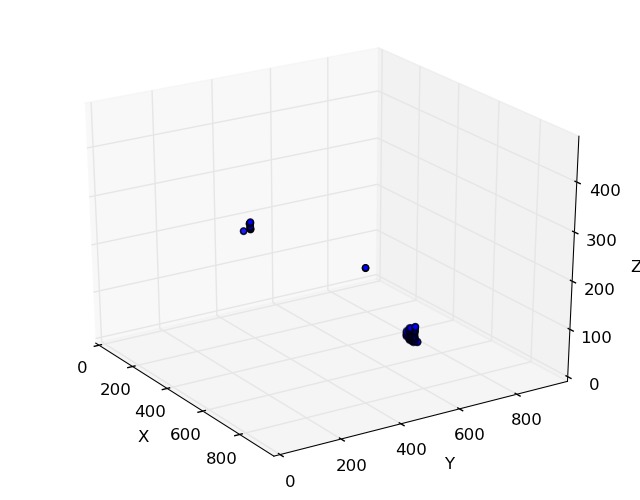}
\includegraphics[width=.23\textwidth]{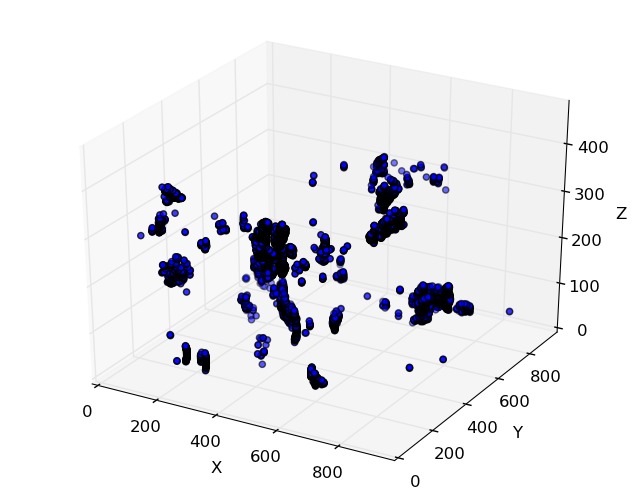}
\includegraphics[width=.23\textwidth]{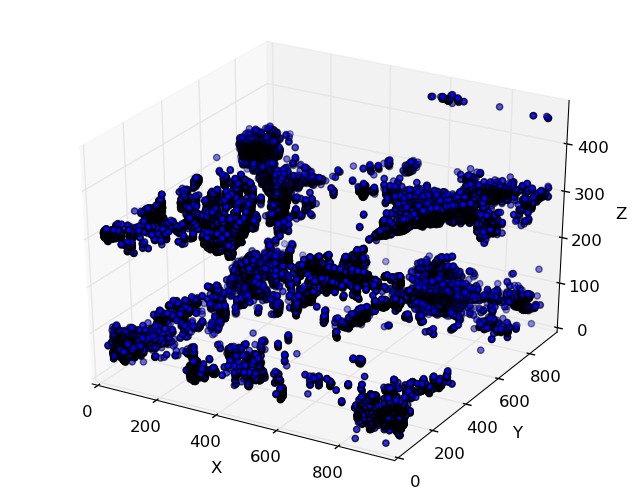}
\includegraphics[width=.23\textwidth]{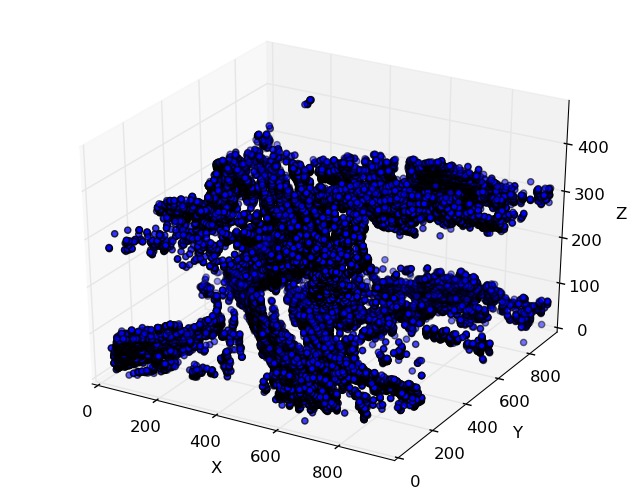}
\caption{Regions with negative potential vorticity are shown from
  start of instability at time 4.5 to a time of 6.2, when the fraction of
  domain with negative potential vorticity peaks. Intermediate times
  shown are 5.1 and 5.7. The volume of domain where potential
  vorticity is negative is highly exaggerated---at time 6.2,
  the region of negative potential vorticity occupies about 0.03\% of
  the total domain.}
\label{fg:pvminloc}
\end{figure}

Figure~\ref{fg:pvminloc} shows regions of negative potential vorticity
at four times starting at time 4.5 when the minimum of potential
vorticity first becomes negative to time 6.2 when the fraction of the
domain where potential vorticity is negative reaches its peak value of
about 0.03\% (see Fig.~\ref{fg:fraction}). The two intermediate times
are 5.1 and 5.7. This figure shows that even though the fraction of
the domain where fronts and unbalanced instabilities occur are small,
they are not localized, but rather well spread out in the domain.

\subsection{Potential Vorticity Snapshots in the Boussinesq System}
\begin{figure}
\centering
\includegraphics[trim=0 0 0 0, clip,
width=0.49\textwidth]{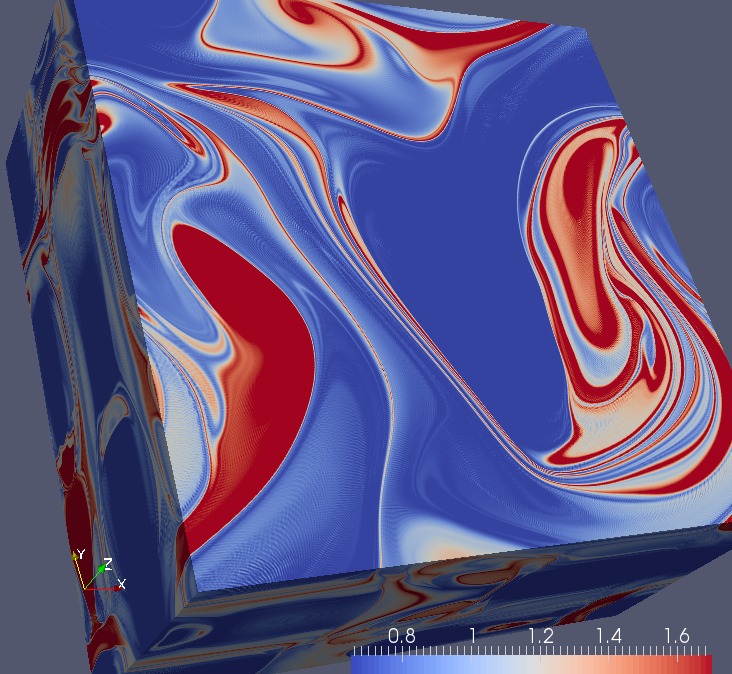} 
\includegraphics[trim=0 0 0 0, clip,
width=0.49\textwidth]{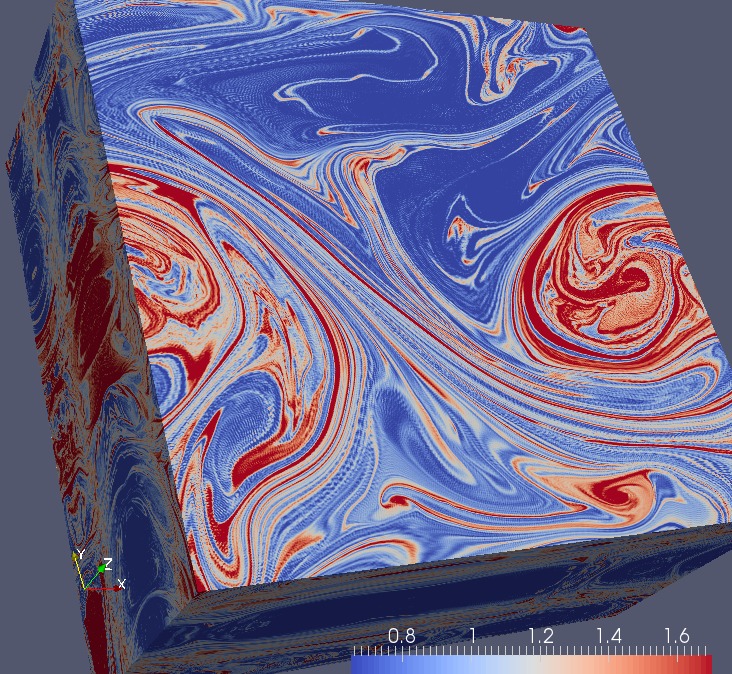}
\includegraphics[trim=0 0 0 0, clip,
width=0.49\textwidth]{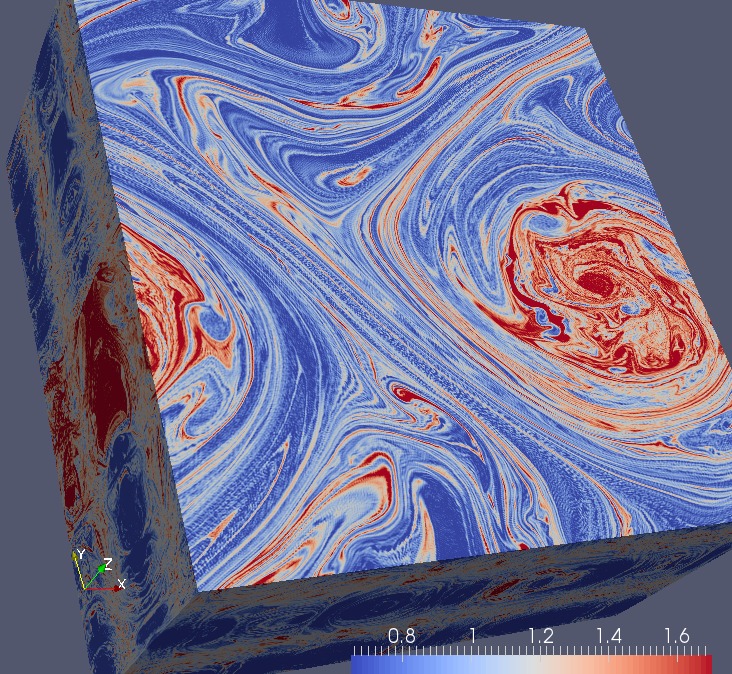}
\includegraphics[trim=0 0 0 0, clip,
width=0.49\textwidth]{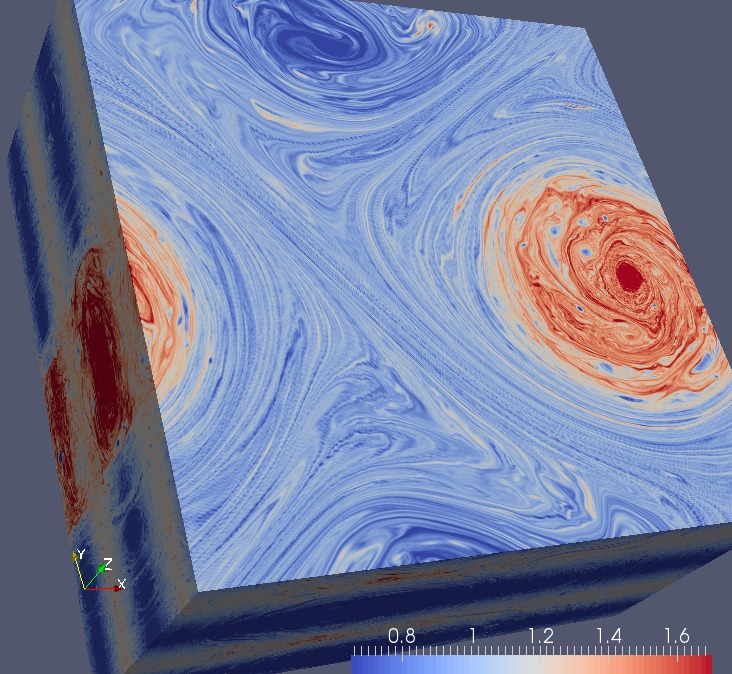}
\caption{Snapshots of potential vorticity in the Boussinesq system. The times
  of the snapshots are close to the time of: top-left: the local peak
  in $\Ro_\omega$ around 4.5; top-right: local minimum of $Ro_\omega$
  and the local minimum of $A-S$ around 6.6; bottom-left: local
  maximum of $\Ro_\omega$ near 8.1; bottom-right: at the late time of 15.}
\label{fg:nbpv}
\end{figure}

Like with the QG system, Fig.~\ref{fg:nbpv} shows four snapshots of
Ertel potential vorticity in the Boussinesq system. 

The times of the
snapshots are chosen to be close to the time of the local peak in
$\Ro_\omega$ around 4.5 in Fig.~\ref{fg:minmax_rv} (top-left), local
minimum of $Ro_\omega$ and the local minimum of $A-S$ around 6.6 in
Figs.~\ref{fg:minmax_rv} and \ref{fg:ams} respectively
(top-right),local maximum of $\Ro_\omega$ near 8.1 in
Fig.~\ref{fg:minmax_rv} (bottom-left), and at the late time of 15
(bottom-right). 

The times of the four snapshots are chosen as follows: The time of the
snapshot in figure 14(a) corresponds to the time of the local peak in
$\Ro_\omega$ in figure 4 near time 4.5. The snapshot in figure 14(b)
is at a time 6.6, a time close when a local minimum of $Ro_\omega$
occurs in figure 4 and a local minimum of $A-S$ occurs in figure
5. That in figure 14(c) is at a time 8.1, around which a local maximum
of $\Ro_\omega$ occurs in figure 4.  Finally, the snapshot in figure
14(d) is at the late time of 15.

The similarity in the phenomenology of the large
scales noted earlier is reflected in the gross similarity of the
large-scale features in Fig.~\ref{fg:nbpv} and \ref{fg:qgpv}. The most
noticeable difference between the two figures, however, is that
small horizontal scales are more energetic in the Boussinesq
system. This is consistent with the phenomenology presented previously
that include the ready formation of fronts and further susceptibility
of these frontal regions to imbalanced instabilities in the Boussinesq
system.

\section{Scaling of Dissipation with Rossby number}
\begin{figure}
\centering
\includegraphics[width=\textwidth]{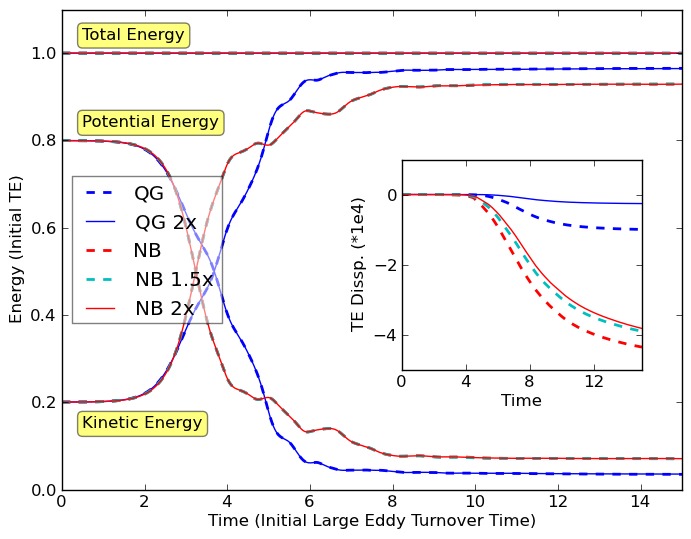}
\caption{Evolution of kinetic, potential and total energy in the
  reference run in the quasi-geostrophic and Boussinesq systems at
  two different resolutions. While early development in the two
  systems are similar (up to about time 5), later development is seen
  to be different with a marked slowdown in the conversion from
  potential energy to kinetic energy in the Boussinesq system. The
  main effect of resolution is seen in the dissipation of total
  energy, as shown in the inset plot. In this inset plot, dissipation
  of total energy is shown in hundredths of a percent of total energy.}
\label{fg:kpt}
\end{figure}

We first consider dissipation in the reference simulation.
The main panel of Fig.~\ref{fg:kpt} shows the evolution of kinetic,
potential, and total energy at two resolutions (1x and 2x) in the QG
system and at three resolutions (1x, 1.5x, and 2x) in the Boussinesq
system for the reference case.  The inset shows the evolution of total
energy on a highly magnified scale.  The initial growth of the
baroclinic instability is seen to be similar in the two systems and at
the different resolutions (main panel). At about four eddy turnover
times, while baroclinic instability has already resulted in the
conversion of a large fraction of the available potential energy into
kinetic energy (main panel), total energy is seen to be very weakly
dissipated (inset).
%
%

In the main panel of Fig.~\ref{fg:kpt}, kinetic and potential energy
evolution is seen to depend only rather weakly on resolution. This is,
however, not the case with total energy or equivalently its
dissipation shown in the inset plot.  The vertical axis of the inset
plot corresponds to dissipation in terms of a hundredth of a percent
of the initial total energy.  Although the level of dissipation is
seen to depend on resolution in both systems, in terms of fractional
change, this dependence is much stronger in the QG system. Indeed,
dissipation in the QG system is verified to approach zero using
Richardson's deferred approach to the limit of infinite resolution. In
the Boussinesq system, dissipation is seen to approach a finite
value. Next we quantify the dependence of this value on Rossby number.

\begin{figure}
\centering
\includegraphics[width=\textwidth]{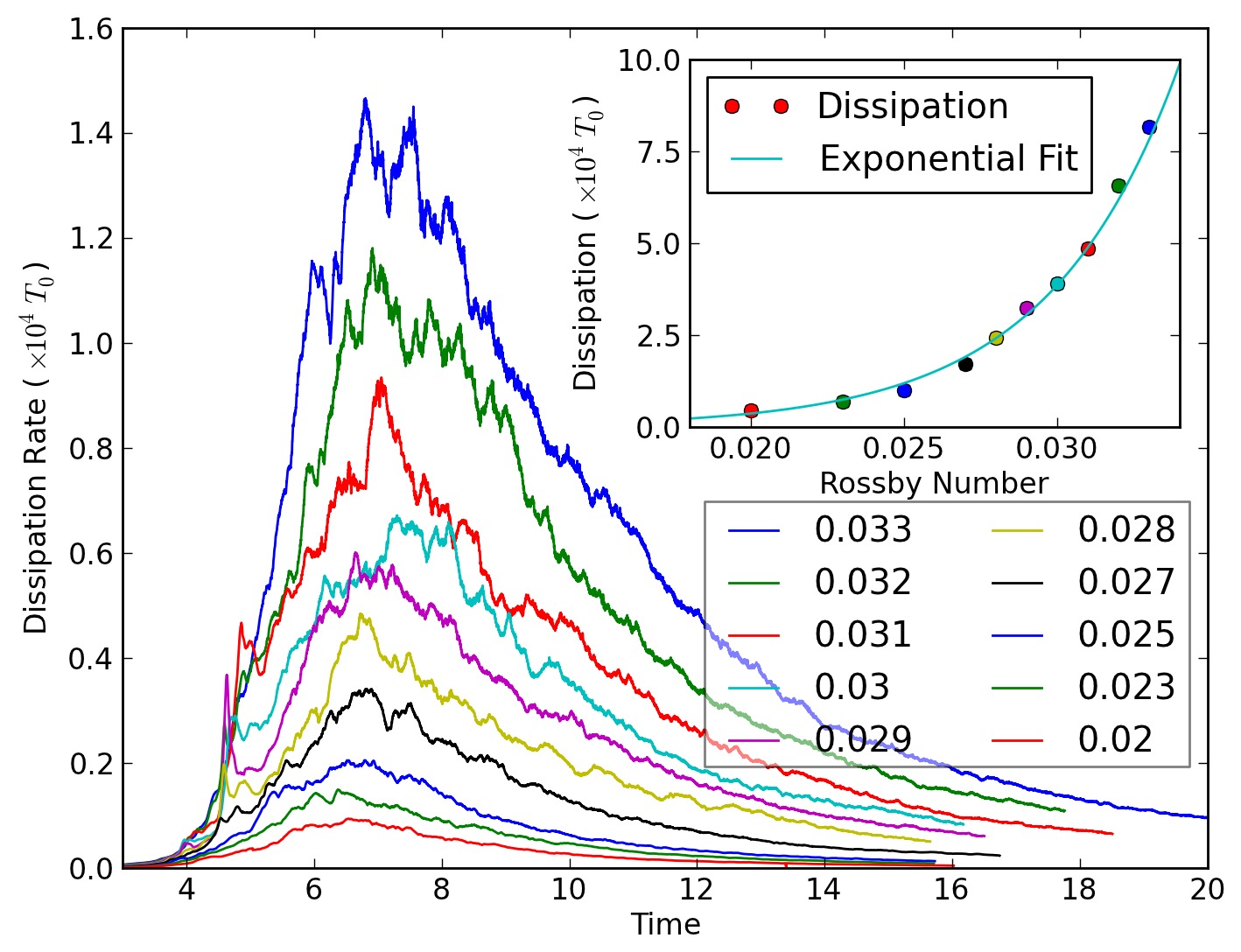}
\caption{Main panel: Dissipation rate as a function of time at various
  Rossby numbers. Corresponding Rossby numbers are shown in the
  legend. Cumulative dissipation at
  time 15.7 is plotted as a
  function of Rossby number using the same colored circles in the
  inset. The continuous line in the inset corresponds to a least-squares
  exponential fit of the dissipation values.}
\label{fg:dssp}
\end{figure}

\begin{figure}
\centering
\includegraphics[width=.49\textwidth]{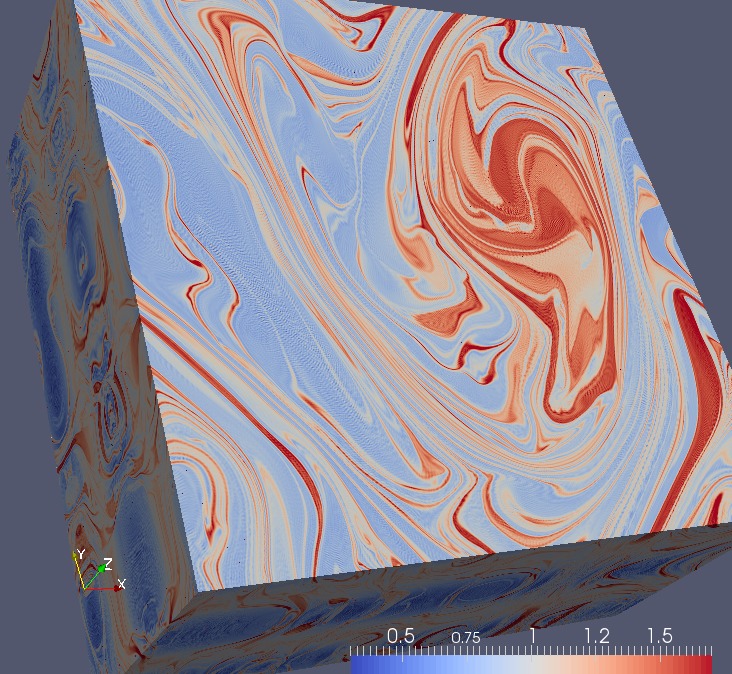}
\includegraphics[width=.49\textwidth]{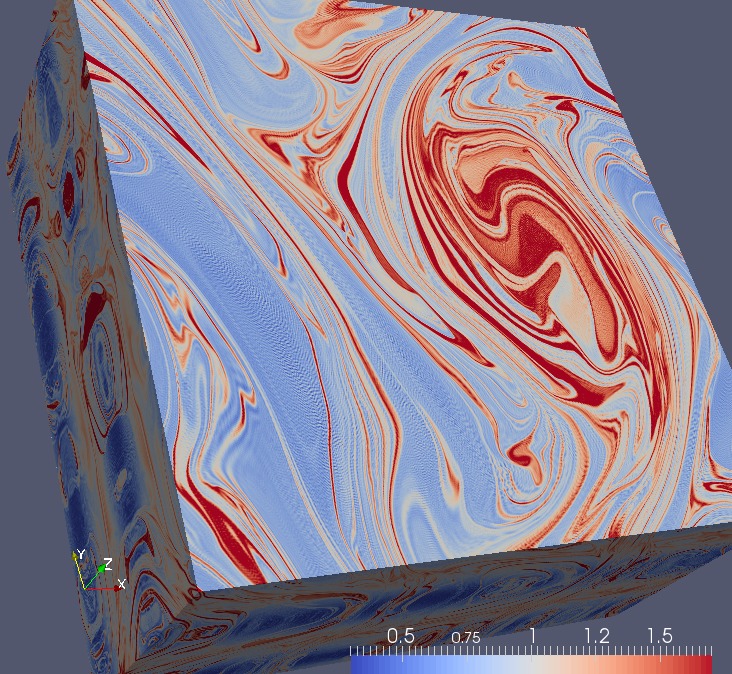}

\includegraphics[width=.49\textwidth]{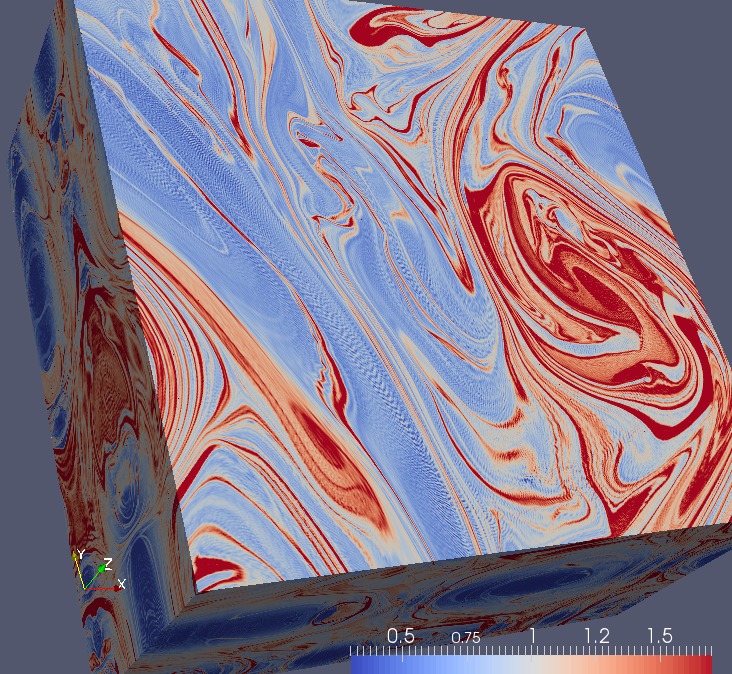}
\includegraphics[width=.49\textwidth]{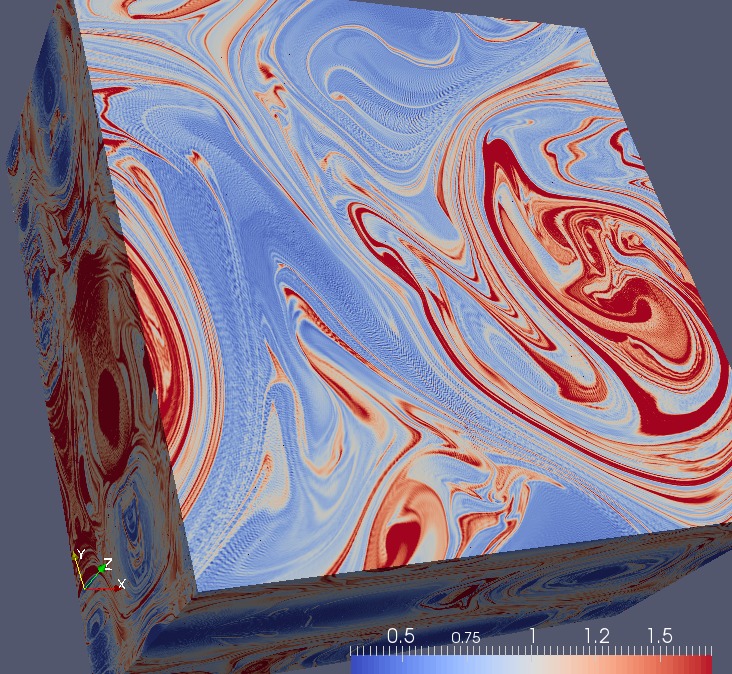}

\includegraphics[width=.49\textwidth]{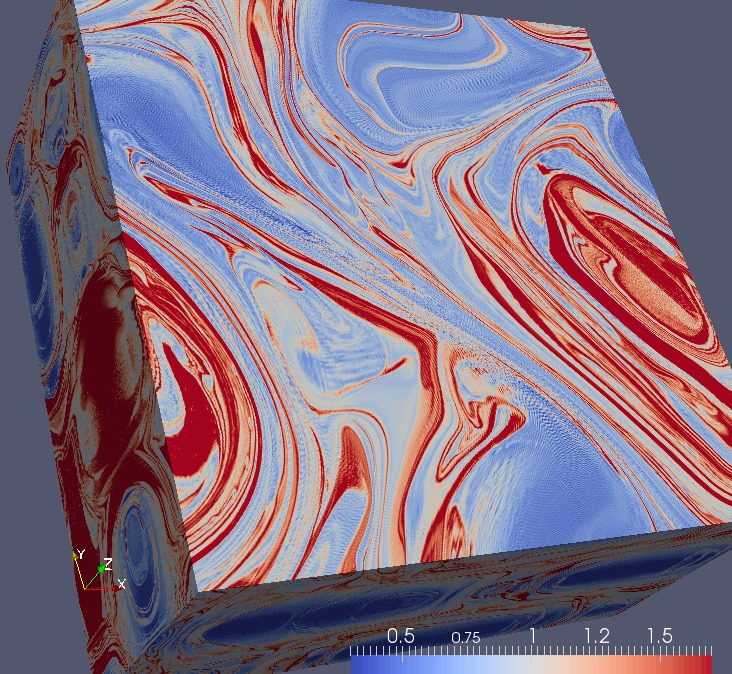}
\includegraphics[width=.49\textwidth]{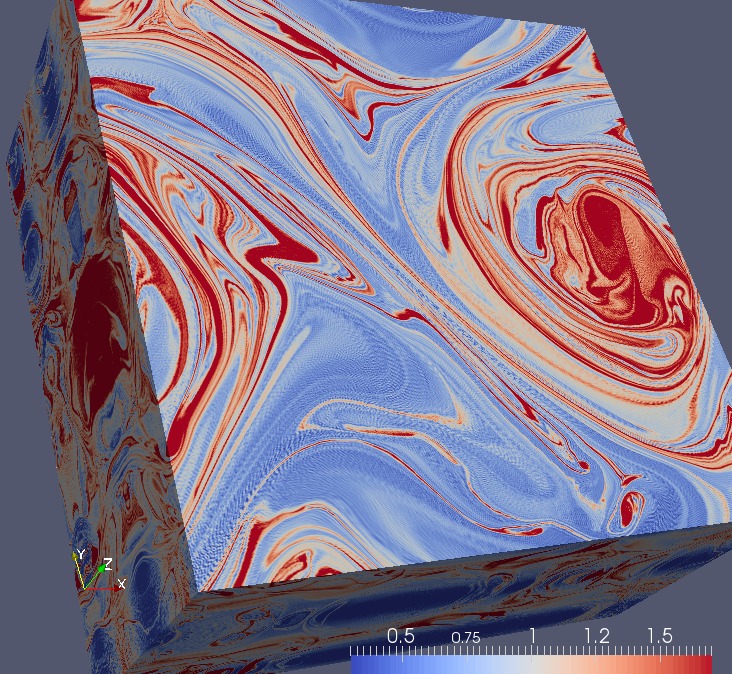}

\caption{Snapshots of the potential vorticity field at time 6 in six of the ten
  Boussinesq cases considered in Fig.~\ref{fg:dssp}. Rossby numbers of 0.02, 0.025,
0.029, 0.03, 0.031, and 0.033 going from top-left to bottom-right. At
this time, dissipation rate is seen to be close to peak values, but
dissipation rate is still rising. Frontogenetic processes are likely
still active.}
\label{fg:pvat6}
\end{figure}

\begin{figure}
\centering
\includegraphics[width=.49\textwidth]{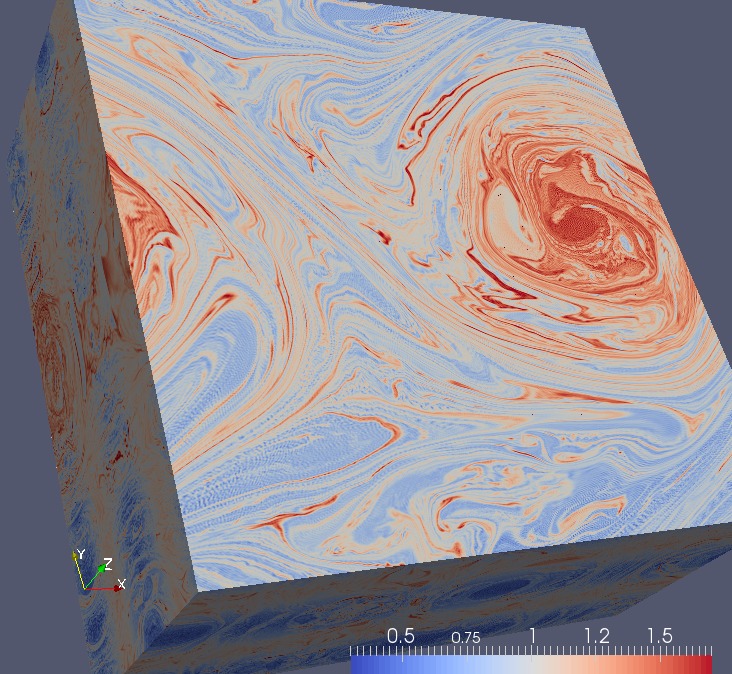}
\includegraphics[width=.49\textwidth]{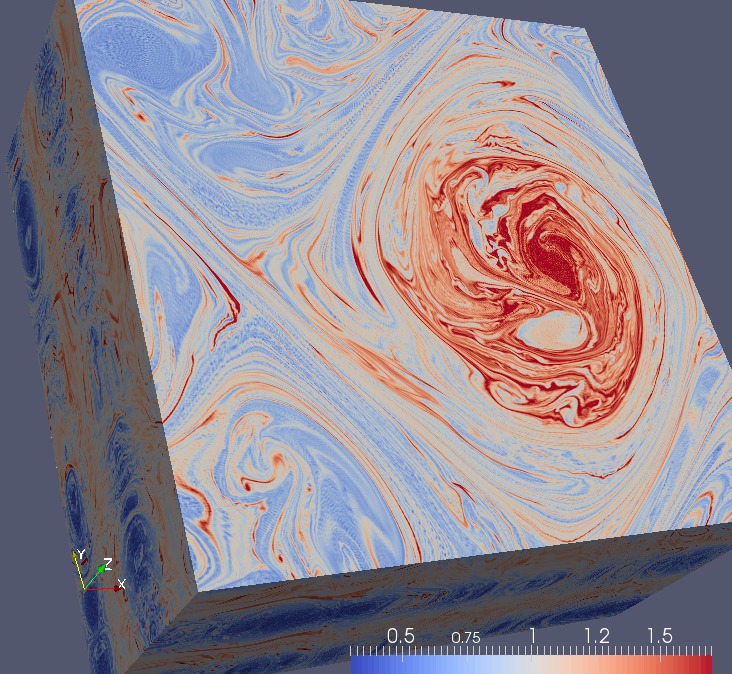}

\includegraphics[width=.49\textwidth]{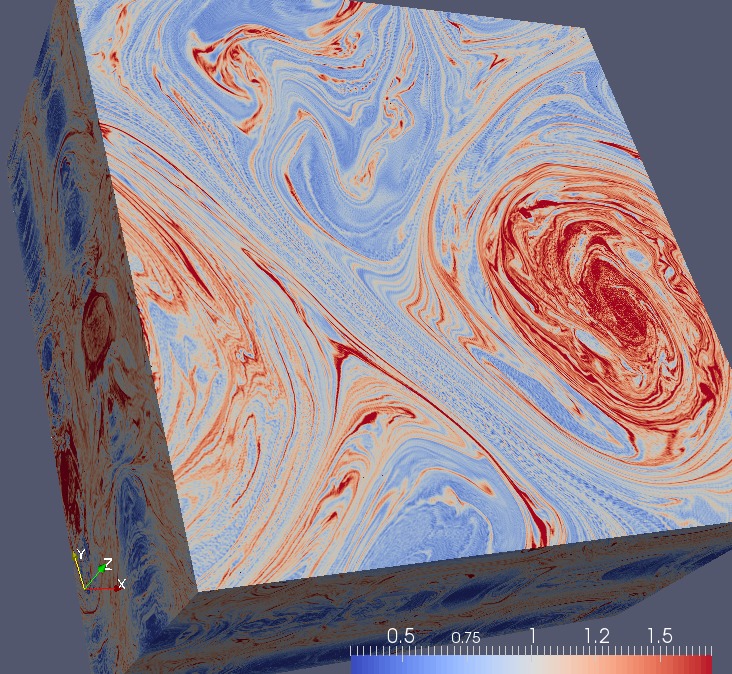}
\includegraphics[width=.49\textwidth]{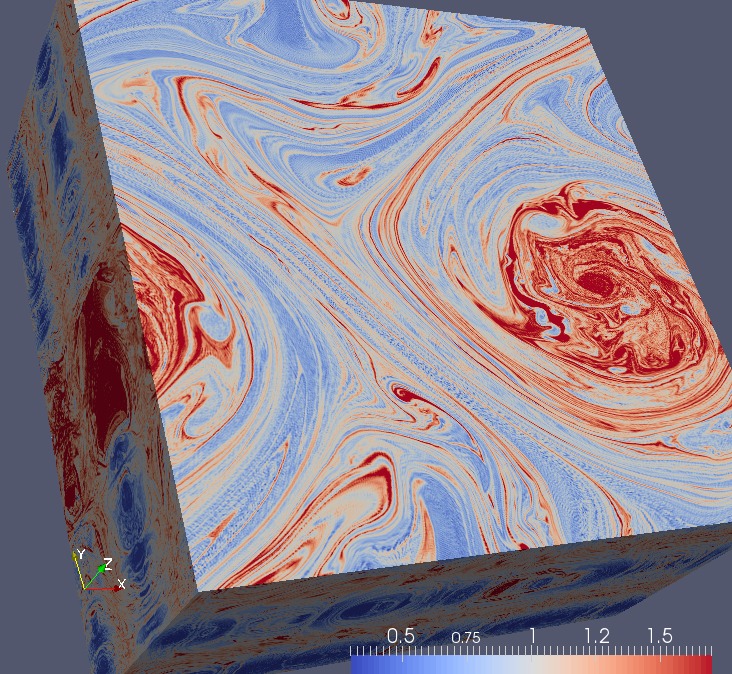}

\includegraphics[width=.49\textwidth]{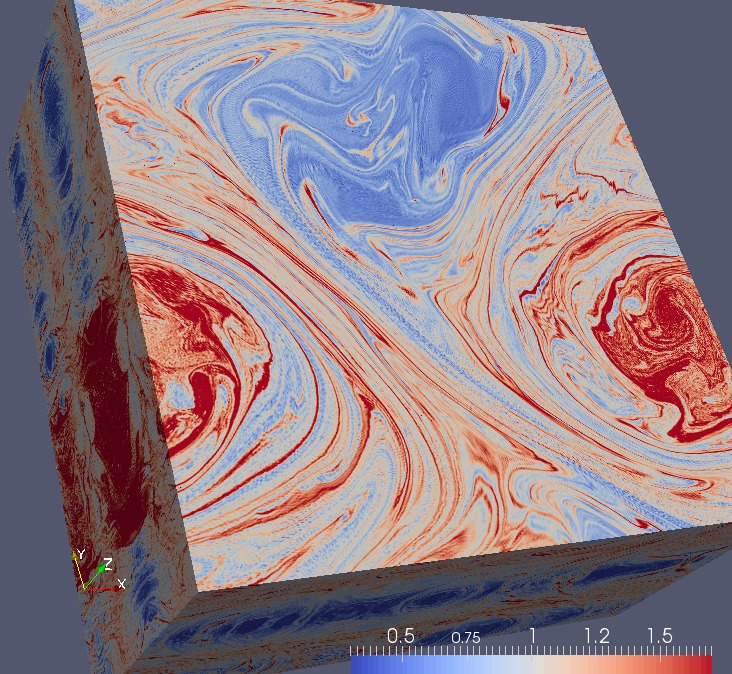}
\includegraphics[width=.49\textwidth]{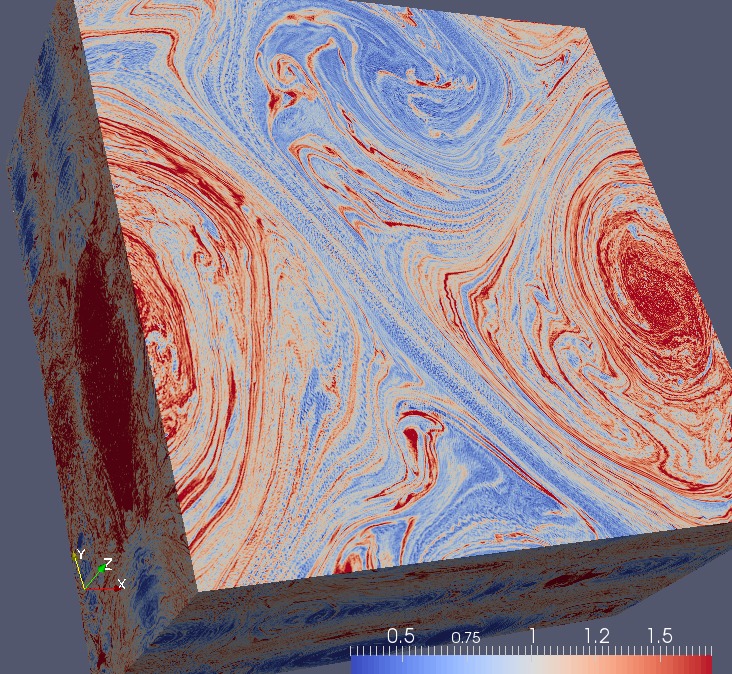}

\caption{Same as if Fig.~\ref{fg:pvat6} but at a later time of 8 when
  dissipation rate is decreasing. Absence or reduced frontogenesis and
increased break-up of fronts by imbalanced secondary instabilities is
the likely reason for qualitative differences from Fig.~\ref{fg:pvat6}.}
\label{fg:pvat8}
\end{figure}

\subsection{Scaling with Rossby Number}
As mentioned previously, one of the main motivations for this work is
to examine the energetic consequences of imbalanced instabilities in
the interior. In the previous section, it was seen that dissipation
of total energy in the quasi-geostrophic approximation displayed the
familiar vanishing behavior whereas that in the Boussinesq system, it
approached a finite value with increasing resolution. In this section,
we examine how dissipation in the Boussinesq system varies with Rossby
number. 

To this end, we varied the velocity-based Rossby number of the initial
condition over a range of values in nine other 'NB 2x' experiments. We note
here, that in order to preserve similarity in the initial evolution
between the quasi-geostrophic and Boussinesq systems (as seen in the
previous section), the ratio of the Froude number to Rossby number was
held fixed at its value (of two) in the reference run. 

The main panel of Fig.~\ref{fg:dssp} shows the time evolution of the
rate of dissipation of total energy in the main sequence of 'NB 2x'
runs. The legend indicates the Rossby number of the corresponding
profiles. (It is hoped that the ambiguity in line type is not a cause
for confusion.) These curves are seen to display a characteristic
asymmetry in that the rise in the dissipation rate between times 4 and
about 7 is steeper than the fall from the peak. As discussed in the
previous section, the steeper rise is due to frontogenesis being
active over that period. Indeed, a slow down in the formation of
fronts is the cause for the peak. 

In the inset plot in Fig.~\ref{fg:dssp}, the total dissipation at time
15.7 is plotted as a function of Rossby number for each of the
experiments. The least squares exponential fit is drawn as a
continuous line and it is seen to be a good fit. In order to rule out
the possibility of numerical artifacts contributing to the exponential
scaling, the 'QG 1x' series of experiments over the range of Rossby
number from 0.02 to 0.03 were analyzed in an identical fashion. While
cumulative dissipation in the Boussinesq system (as seen in the inset
plot Fig.~\ref{fg:dssp}) varied by about a factor of 10 over this
range of Rossby numbers, the variation in the QG system was about
0.1\%: In the inset plot, QG dissipation as a function of Rossby
number would be visually indistinguishable from a horizontal
line. These comparisons suggest that numerical artifacts contribute
minimally to the exponential increase of dissipation with Rossby
number in the Boussinesq system.

Figures~\ref{fg:pvat6} and \ref{fg:pvat8} show snapshots of potential
vorticity at times close to 6 and 8 respectively in six of the ten
cases considered in Fig.~\ref{fg:dssp} (Rossby numbers of 0.02, 0.025,
0.029, 0.03, 0.031, and 0.033 going from top-left to
bottom-right). Both these times are times when dissipation rate is
seen to be high and in fact close to peak values in
Fig.~\ref{fg:dssp}. On this count it would seem surprising that the
fields are qualitatively quite different at the two times. However, at
time 6 dissipation rate is still rising unlike at time 8 when
dissipation rate is on its way down. Indeed, one of the main difference in
the potential vorticity fields at these two times is that the elongated filamentary
structures are more coherent at time 6 than they are at
time 8. A plausible explanation of this difference is that the faster
rise in dissipation rate in Fig.~\ref{fg:dssp} is in part due to
the forward cascade of energy associated with the formation of
fronts. Note that this is a statistical statement about the overall state of
the flow considering that the timescale for the formation of fronts is
likely much shorter than a large-scale eddy turnover time. On the
other hand, a slower fall-off in the dissipation rate past a time of
about 7 is likely due to an absence or at least a reduced rate of
formation of fronts; the main processes that underlie dissipation
during this phase are the secondary imbalanced instabilities. We will
present further evidence to support this explanation in the next
section when we discuss nonlinear spectral flux of energy across scale.

\section{Spectral Scaling and Spectral Flux of Energy Across Scale}
\label{sec:spectra}
\begin{figure}
\centering
\includegraphics[width=\textwidth]{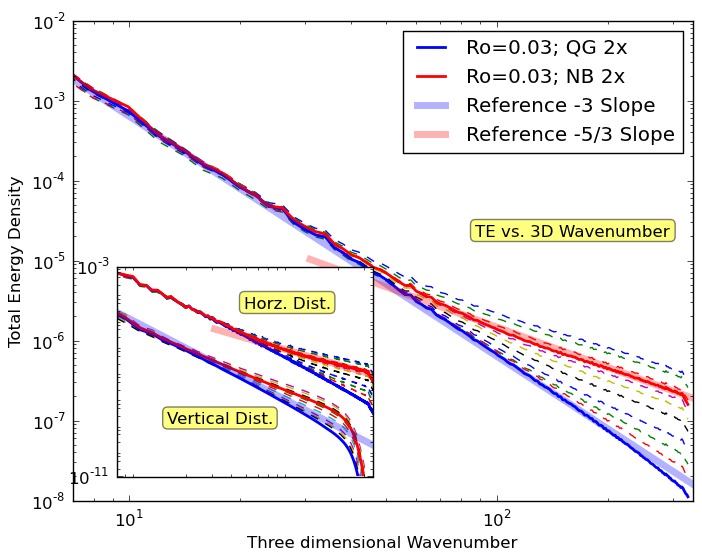}
\caption{Total energy spectra in the QG and Boussinesq systems with
  2x resolution. The cases considered in detail in previous
  diagnostics ($\Ro=0.03$) are shown in heavy lines: blue: QG, red:
  Boussinesq. The dashed lines correspond to Boussinesq cases with
  $\Ro$ ranging from 0.02 to 0.033.
  In the main panel, spectra are plotted against the three dimensional
  wavenumber where the vertical is scaled by $N/f$. In the inset,
  spectra are plotted against the horizontal 
  wavenumber for the top set of curves and against the vertical wavenumber
  (scaled by $N/f$) for the lower set of curves. Spectra are averaged
  between times 5 and 13
  A shallowing of the spectra from the $-3$ quasi-geostrophic scaling 
 is seen over an increasing range of small scales at $\Ro$ is
 increased.}
\label{fg:te_spec}
\end{figure}

We now examine spectral scaling of total energy as a tool to diagnose
imbalance.  In the main panel of Fig.~\ref{fg:te_spec}, spectra of
total energy are plotted as a function of the three dimensional
wavenumber after scaling the vertical by $N/f$. In the inset plots,
distribution of total energy with horizontal and scaled vertical
wavenumbers are plotted; the vertical spectra in the inset are offset
by a decade and a half for ease of presentation. The horizontal axis
in the inset plot spans the same range of scales as in the main panel
and is omitted to reduce clutter.  The QG and Boussinesq cases
compared in previous diagnostics ($\Ro=0.03$) are shown in heavy
lines, with the QG case shown in blue, and the Boussinesq case in red.
The dashed lines correspond to series of Boussinesq cases with $\Ro$
ranging from 0.02 to 0.033.  Recalling that the energy spectrum in the
balanced regime of forward potential enstrophy cascade scales as $-3$,
this slope is shown for reference. It is seen in this plot that the
intermediate range of scales (wavenumber less than about 30) in all
the Boussinesq cases and most of the scales in the QG 2x case display
the balanced $-3$ scaling.  The slightly steeper fall-off at small
scales in QG 2x is not unusual and has been previously noted in many
instances; it is thought to be related in part to the existence of
coherent structures.

In the Boussinesq system, it is seen that as $\Ro$ increases, there is
an increasing range of small scales where the spectrum displays a
shallower fall off.  Processes that lead to deviations from a balanced
regime and which have the potential to lead to a forward cascade of
energy are expected to lead to shallower spectral slopes \citep[e.g.,
see][]{bartello95}. However, the exact dynamics associated with this
shallower fall-off is a topic of debate and an area of on-going
research. For example, it has been suggested that when rotational
control is reduced at small scales, there is likely to be a regime of
anisotropic stratified turbulence before a regime of more isotropic
turbulence is realized at yet smaller scales. For example, Lindborg
and co-workers \citep{lindborg2006energy, riley2008stratified} suggest
that the kinetic and potential energy spectra each scale as
$k_h^{-5/3}$ in the horizontal and $k_z^{-3}$ in the vertical in the
regime of stratified turbulence where energy cascades forward. While
their scaling in the horizontal is the same as in the regime of
isotropic three-dimensional turbulence where energy cascades forward
as well, anisotropy of dynamics in the stratified regime leads to a
steeper fall-off of the energy spectrum in the vertical. For this
reason, a reference $-5/3$ slope is indicated on the three-dimensional
wavenumber distribution and horizontal wavenumber distribution
plots. Indeed, at small scales, the spectra of total energy against
the three-dimensional wavenumber and the horizontal wavenumber is seen
to match this slope over an increasing range of scales for increasing
Rossby number. We expect that with yet higher resolutions, the lower
Rossby number cases will also display shallower spectra over a larger range
of scales than seen with the present resolution.

\subsection{Vertical Shear Spectrum and Stratified Turbulence}
The spectral fall-off of the total energy spectrum with vertical
wavenumber is seen to be shallower than the $-3$ spectral fall-off of
stratified turbulence \citep{lindborg2006energy}. At first glance, this
suggests that a regime of stratified turbulence that is presumably fed
by imbalanced instabilities is not clearly realized in these
simulations. However, as mentioned earlier, the spectra in
Fig.~\ref{fg:te_spec} are averaged between times 5 and 13 which spans
different regimes from the point of view of stratified turbulence.  

\begin{figure}
\centering
\includegraphics[width=\textwidth]{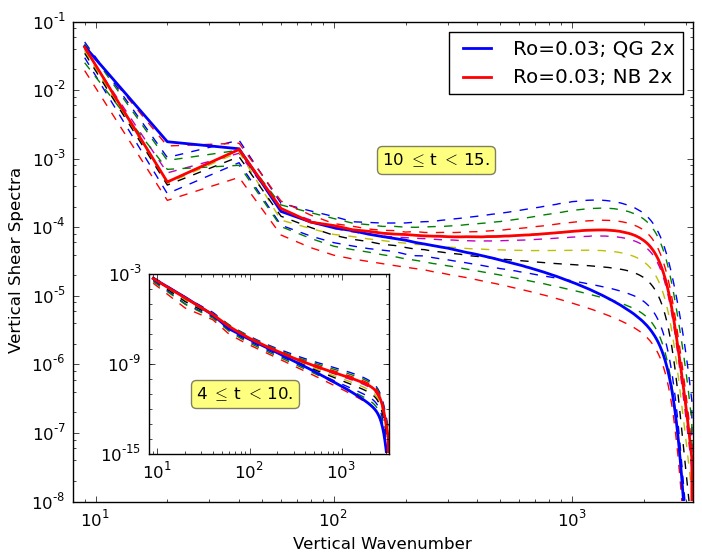}
\caption{Spectrum of vertical shear of horizontal velocity as a
  function of vertical wavenumber for the main sequence of 2x cases
  considered. Main panel shows the average over late times whereas the
  inset shows the average over intermediate times. A minimum in this
  spectrum, seen here at late times for the larger of the Rossby
  numbers considered, is indicative of a transition from anisotropic
  stratified turbulence to more isotropic turbulence. A similarity of
  this figure to Fig. 1 of \cite{gargett1981composite} which is a
  composite based on in-situ measurements is noted.}
\label{fg:vertshear}
\end{figure}

We examine further if indeed the simulations are entering a regime of
stratified turbulence. To this end, we note that from
Fig.~\ref{fg:rms_Ro_Fr} (where the rms value of $\Fr_\omega$ becomes
small at late times) and from the inset plot in
Fig.~\ref{fg:minmax_rv} (where the peak value of $\Fr_\omega$ is
smallest, again at late times), stratified turbulence, if realized, is
most likely to occur at late times.  Consequently,
Fig.~\ref{fg:vertshear} shows the spectral distribution of the
vertical shear of horizontal velocity as a function of vertical
wavenumber in all the cases considered. A minimum in this spectrum,
occurring in the vicinity of the wavenumber corresponding to the
Ozmidov scale, is interpreted as indicative of a transition between
those scales of motion that are strongly influenced by stratification
and those that are not \citep{riley2008stratified}. 

A minimum is indeed seen in these spectra in Fig.~\ref{fg:vertshear}
at late times for the higher of the Rossby numbers considered and
occurs between wavenumbers 100 and 200.  Further, the spectra in
Fig.~\ref{fg:vertshear} are qualitatively similar to those in Fig. 1
of \cite{gargett1981composite} which is a composite based on in-situ
measurements. (This latter figure is reproduced and discussed in
\cite{riley2008stratified}.)  Nevertheless, further detailed
investigation is necessary to assess the relevance of these
simulations to the in-situ measurements of
\cite{gargett1981composite}.

It is also clear that a minimum is not seen in Fig.~\ref{fg:vertshear}
for the smaller of the Rossby numbers considered. This transition as a
function of Rossby number from having a minimum to not having one is
rather continuous, with the minimum becoming shallow before
disappearing.  Which leads us to suggest that the question is not one
of whether or not stratified turbulence spans the regimes of
anisotropic rotating-stratified turbulence at large scales and more
isotropic turbulence at small scales, but one of how important it is
in the particular case or parameter regime of interest.

\begin{figure}
\centering
\includegraphics[width=\textwidth]{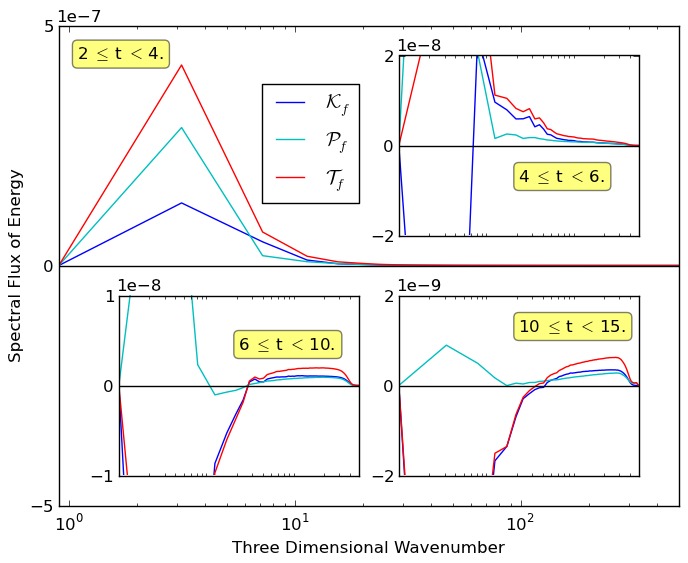}
\caption{Spectral flux of kinetic (blue), available potential (cyan),
  and total (red) energy in the Boussinesq system with
  2x resolution for the reference case ($\Ro=0.03$) averaged over four
  different periods} 
\label{fg:nbflux_spec}
\end{figure}

\begin{figure}
\centering
\includegraphics[width=\textwidth]{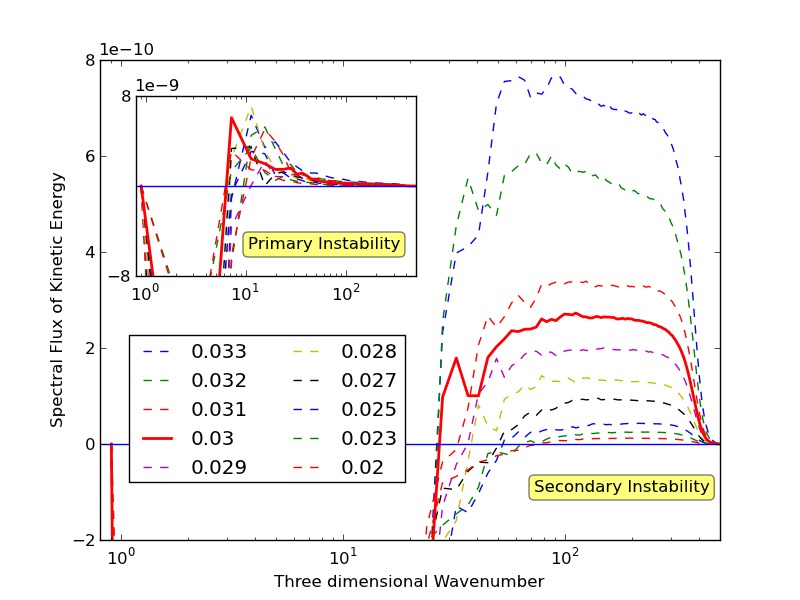}
\caption{Spectral flux of energy  in the Boussinesq system with
  2x resolution. The $\Ro=0.03$ case considered in detail in previous
  diagnostics is shown in heavy red line; other cases with
  $\Ro$ ranging from 0.02 to 0.033 are shown as thinner lines.
  In the main panel, spectral flux of kinetic energy is plotted
  against the three dimensional wavenumber averaged over 13 equi-spaced snapshots
  between times 6 and 10, a period when dissipation is high. In the inset
  the average of spectral flux of kinetic energy over 6 snapshots
  between times 4 and 6 is plotted.}
\label{fg:flux_spec}
\end{figure}

\subsection{Spectral Flux of Energy Across Scale}
We first present an overall picture of the flow evolution in terms of 
the spectral flux of kinetic energy (KE), available potential energy
(APE) and total energy (TE) across scale. To this end,
Fig.~\ref{fg:nbflux_spec} shows the spectral flux across scale of
KE $\sK_\str$ (blue), that of APE
$\sP_\str$ (cyan), and that of TE $\sT_\str$ as a function of the
three-dimensional wavenumber (see Section~\ref{sec:energy} and details
around (\ref{eq:transfer})). Averages over four time periods are
shown. 

Between times 2 and 4 the hydrostatic geostrophic baroclinic
instability is in its nonlinear phase mesoscale eddies are being
formed. In this phase, the expected forward cascade of APE is clearly
seen. As for the spectral flux of KE, two processes are at
play. First, note that that initial KE is concentrated at a scale
close to the domain size. When mesoscale eddies form this energy is
essential moves downscale. And then there is the conversion of APE to
KE through baroclinic instability of the initial zonal flow to further
energize the mesoscale eddies. The net effect is seen to be a
downscale cascade of KE. With both KE and APE cascading downscale, so
does TE.

After mesoscale eddies are formed, between times 4 and 6, the first of
the above processes (KE transfer from domain scale to mesoscale) has terminated
and at the large scales an inverse cascade of KE is seen to be
established. At these scales, the forward cascade of APE continues
over this time period, leading to a net forward cascade of TE.  

At wavenumbers greater than about eight, forward cascade of both KE
and APE (and consequently forward cascade of TE) is seen. Recall that
mesoscale shear and strain driven frontogenetic processes are most
active over this time period (see Section~\ref{sec:front}) and explain
the forward cascade of APE. Indeed, the ageostrophic secondary
circulation associated with fronts and the positive feedback process
mentioned in Section~\ref{sec:front} are likely responsible for the
forward cascade of KE seen.

In the third time period shown ($6\leq t < 10$), in the range of
scales associated with the hydrostatic geostrophic baroclinic
instability, the forward cascade of APE is winding down whereas the
inverse cascade of KE continues leading to an inverse cascade of
TE. Over this time period, a robust forward cascade of KE, APE, and TE
is seen to be established. Indeed, over this period, with
frontogenetic processes winding down, while at the same time
conditions for inertial and symmetric instabilities are satisfied, the
forward cascade is a result of baroclinic and other 
instabilities that are imbalanced. Here, we note that some of these
instabilities, e.g., fronts as seen earlier, can have very disparate
characteristic dimensions rendering spherically averaged wavenumber a
poor indicator of such characteristic scales of instability---but
hopefully still useful qualitatively. 

The characteristics of the spectral flux of energy over the final time
period considered is seen to be similar to that in the 
previous time period, except that things are winding down. (Note the
different scales for the $y$ axis in the third and fourth periods.)

\subsection{Nonlinear Spectral Flux of Kinetic Energy as a Function of
  Rossby number}
Figure~\ref{fg:flux_spec} shows the spectral flux of KE as a
function of the three-dimensional wavenumber for the main sequence of
Boussinesq cases considered at 2x resolution. In the main panel, this
spectrum is averaged between times 6 and 10. Note that this corresponds
to the third time period considered in the previous subsection and
when a robust forward cascade was seen. For brevity we call this
forward cascade as being caused by a set of secondary imbalanced
instabilities. In the inset an average between times 4 and 6---the
second time period of the previous subsection---when frontogenesis is
most active is shown. A few observations are in order here. First, the
scale of the zero-crossing point of the spectral flux of KE ($\sK_f$) is
much smaller in the front formation phase and is likely related to the
scale of the primary hydrostatic geostrophic baroclinic
instability. This is also consistent with mesoscale shear and strain
driving frontogenesis. The zero-crossing scale in the secondary
instability phase is smaller, but the isotropic spectrum is a poor
indicator of the characteristic scale of instability, as discussed earlier.
A curiosity of this figure is that whereas the
magnitude of the flux in the 'front formation' phase depends 
weakly on the Rossby number, this dependence is strong in the
secondary instability phase. Indeed, since the forward cascade of
kinetic energy leads in part to dissipation we expect that the
dependence of the magnitude of the forward cascade flux in the main
panel is exponential as well, although noisy.

\begin{figure}
\centering
\includegraphics[width=\textwidth]{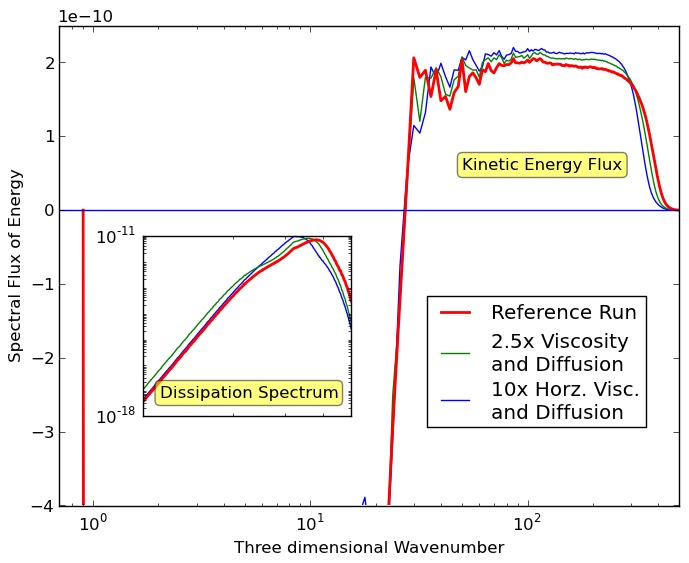}
\caption{Change of spectral flux of energy and dissipation spectrum
  with increases in viscosity and dissipation. In addition to the
  reference case, two cases with increased viscosity and dissipation
  are shown. In the first case, coefficients of both horizontal and
  vertical hyperviscosity and hyperdiffusion are increased by a factor
  of 2.5. In the second case, only the coefficients of horizontal
  hyperviscosity and hyperdiffusion are increased by a factor of
  10. The dependence of both the spectral flux of energy and the
  spectrum of dissipation is seen to be weakly dependent on the
  hyperviscous and hyperdiffusion coefficients.}
\label{fg:dssp_verify}
\end{figure}

Figure.~\ref{fg:dssp_verify} shows the robustness of the spectral flux of
energy---a third order quantity---to changes in the coefficient of
hyperviscosity and hyperdiffusivity. Here the reference simulation has
been repeated two more times: In the first instance coefficients of
both horizontal and vertical hyperviscosity and hyperdiffusion are
increased by a factor of 2.5. In the second instance, only the
coefficients of horizontal hyperviscosity and hyperdiffusion are
increased by a factor of 10.

\section{Discussion and Conclusions}
We consider a regime of rotating stratified flow that is of direct
relevance to circulation in the oceans and in atmospheres---a regime
characterized by small large-scale Rossby and Froude numbers and a
small vertical to horizontal aspect ratio. In this regime, it is well
established that quasi-geostrophic theory provides a good description
of the {\em large} scales. An important consequence of the theory is a
confinement of energy to large scales---scales where dissipation is
ineffective. Furthermore, the steep $-3$ scaling of the energy
spectrum due to the forward cascade of potential enstrophy predicted
by quasi-geostrophic theory can be viewed as effectively precluding
deviations from quasi-geostrophy. In such a regime, we consider the
question of how a flow forced at large scales might
equilibrate. Whereas a large body of recent research has demonstrated
the importance of surface related processes in the equilibration, our
interest in this article has to do with interior processes. 

We use the framework of baroclinic instability---one of the main
building blocks of geophysical fluid dynamics---to setup and study the
problem.  In this context, we note that baroclinic instability has
typically been studied in three different setups \cite[e.g.,
see][]{vallis06}.  In the first setup---the Phillips problem---the
fluid is discretized into layers in the vertical leading to
discontinuous stratification and the instability is driven by interior
gradients of potential vorticity.  In the second setup---the Eady
problem---the instability is driven by surface buoyancy gradients at
two bounding horizontal surfaces. In the third setup---the Charney
problem---the instability is driven by a combination of interior
potential vorticity gradients and boundary buoyancy gradient at a
horizontal surface.  Since the interaction of energy-bearing balanced
scales---oceanic mesoscale eddies---with boundaries such as the
oceanic surface layer has the potential to support balanced-unbalanced
energy interactions, the Eady problem or the Charney problem---in the
framework of either the nonhydrostatic Boussinesq equations or the
hydrostatic primitive equations---has been the preferred setup to
investigate such interactions \citep[e.g.,
see][]{molemaker2010balanced, thomas-tandon08}. To investigate the
more direct local route of balanced-unbalanced interactions in the
oceanic interior where the bulk of balanced energy resides, a setup in
which baroclinic instability is driven by interior gradients of
potential vorticity, as in the Phillip's problem, is more appropriate.
Furthermore, with the possibility of overturning instabilities playing
a role in balanced-imbalanced interactions, a model with continuous
stratification is more appropriate.  To this end, we considered a
setup with continuous stratification and in a periodic geometry---a
setup that explicitly eliminates distinguished horizontal
surfaces. This setup may be thought of as an extension of the Phillips
model to one with continuous stratification, yet with no bounding
horizontal surfaces,  or equivalently as an extension of the Charney
model to a periodic domain.

We then studied the nonlinear evolution of baroclinically unstable
initial conditions that comprised of a geostrophically and
hydrostatically balanced unidirectional base flow solution and a set
of small perturbations, each of which was a balanced solution of the
inviscid nonlinear equations. Initially energy was confined to a scale
close to the size of the domain. This energy was then transferred
downscale starting first with the hydrostatic geostrophic baroclinic
instability. The resulting mesoscale shear and strain led to
frontogenesis that then preconditioned the flow to be susceptible to
unbalanced inertial and symmetric instabilities. These instabilities
led to a forward cascade of energy that terminated in energy
dissipation by parameterized hyperviscous and hyperdiffusive
dissipation. The total dissipation over the period of the instability
was less than about a tenth of a percent of the total energy and we
found an exponential dependence of dissipation on Rossby number of the
base flow.

As control, we carried out companion integrations or twin
experiments using the quasi-geostrophic equations. A number of
flow characteristics of the balanced quasi-geostrophic evolution and the
evolution of the nonhydrostatic Boussinesq system were compared. There
were similarities in gross characteristics of the evolution and
significant differences in certain other detailed characteristics. We
interpreted these differences in terms of previous theoretical
analysis of instabilities. In the companion quasi-geostrophic
experiments, dissipation was seen to be vanishingly small and no
exponential dependence on the Rossby number of the base state was seen.

Given the fundamental nature of the dynamics involved, it is possible
that the exponential scaling of dissipation with Rossby number we find
will hold more generally than for the specific setup we consider. In
particular, it would not be surprising if a similar scaling was
realized in setups with surface layers or boundary layers. A natural
question regarding this dissipation pathway, however, is as to how
important it is to the energy budget of ocean circulation
\cite[e.g.,][]{ferrari2008ocean}. We recall that in the present setup,
we have chosen a constant background stratification $N_o^2$. On the
one hand, from previous studies of geostrophic turbulence
\citep[e.g.][]{fu1980nonlinear} we know that in a more realistically
stratified ocean, characteristics of the inverse cascade of energy are
significantly different from that in which stratification is
constant. On the other hand, recent studies such as that of
\citet{plougonven2009nonlinear} show that nonlinear saturation of an
imbalanced instability such as inertial instability even in a simple
setting such as barotropic shear in a constantly stratified
environment results in complicated vertical structure. Since
dissipation would likely depend on interactions between processes such
as these, estimating the importance of this energy pathway would have
to depend on further numerical studies with realistic stratification
and shear.

In other findings, we considered the spectrum of total energy as a
function of three-dimensional wavenumber and found a break in
the spectrum at small scales for the larger of the Rossby numbers
considered. While the intermediate scales in the Boussinesq system and
most scales in the quasi-geostrophic system scaled as $k^{-3}$, the
small scales in the Boussinesq system for the larger of the Rossby
numbers considered scaled approximately as $k^{-5/3}$. We also considered
the vertical shear spectrum as a function of vertical scale and found a
minimum in this spectrum at late times for the larger of Rossby
numbers considered. The distribution we found is somewhat reminiscent
of Fig. 1 of \cite{gargett1981composite} which is a composite based on
in-situ measurements and suggests that at late times and in the larger
of the Rossby numbers considered, the simulations may be entering the
regime of stratified turbulence before transitioning into a more
isotropic form.

While the exponential dependence of the growth rate of unbalanced
instabilities on Rossby number is well recognized \citep[e.g.,
see][and references therein]{molemaker05, vanneste2013balance}, we
think that the exponential scaling of dissipation on Rossby number of
the base flow we find is a further useful step in better
characterizing balanced-imbalanced interactions. Indeed, to the best
of our knowledge, this study is the first of its kind in this
respect. We speculate that a consequence of this finding could be the
introduction of a dissipative component to the parameterization of
baroclinic instability: Ocean modeling has undergone vast improvements
in the last half a century and one of the key improvements has been
the Gent-McWilliams parameterization \citep{gent1990isopycnal} of
baroclinic instability in models that do not resolve the instability
scale.  This parameterization consists of a skew flux that arises from
an inviscid consideration of the slumping of isopycnals that is
characteristic of the instability.  However, preliminary analysis of
simulations presented in this study suggest that a developing
baroclinic instability can lead to secondary instabilities on a range
of scales and that some of these instabilities can cascade energy
forward to unbalanced scales. This suggests that a secondary aspect of
(geostrophic hydrostatic) baroclinic instability is the ultimate
dissipation of energy cascaded forward by some of the secondary
instabilities---a feature that could perhaps be usefully
parameterized. In keeping with advances in the representation of
various unresolved processes, such a modification of the
parameterization of baroclinic instability would likely result in
shifting a component of dissipation represented by the (ubiquitous)
homogeneous eddy viscosity to a more intermittent, process-related
representation.

Investigations of spontaneous generation of imbalance is an area of
extensive and on-going research (e.g. see special collection of the
Journal of the Atmospheric Sciences on 'Spontaneous Imbalance', May,
2009, or, e.g., \citep{plougonven2007inertia,
  danioux2012spontaneous} for studies in the specific context of an
unstable baroclinic wave).  On the face of it, it may seem that the
present work is not related to this area of research. This is because
while the base flow we considered was an exact and balanced solution,
we additionally introduced small perturbations in the initial
conditions. While each of these perturbations, individually, was an
exact solution as well, their superposition on the base flow rendered
the initial condition to not be a solution of the nonlinear equations and 
further likely imbalanced. Therefore it is possible that all
dissipation that ensued is traceable to imbalance in the initial
conditions. In this sense, the imbalance observed in the context of
the simulations considered in this article are likely not
spontaneously generated. On the other hand, it is possible these
results have something to say about the nature of spontaneous
generation of imbalance. To see how this may be possible, we first
hypothesize that, given other things are the same, imbalance in the
initial conditions considered in this article is directly proportional
to the Rossby number of the base flow. If this were true (we think
so), then the exponential dependence of dissipation demonstrated in
this article would suggest that dissipation of balanced mesoscale
energy by processes related to spontaneous generation of imbalance is
likely to be negligibly small.\\

This research was supported by the Laboratory Directed Research and
Development (LDRD) program at Los Alamos National Laboratory (project
number 20110150ER). Computational resources were provided by
Institutional Computing at the Los Alamos National Laboratory. Thanks
to Mark Taylor both for sharing his spectral code.  Brief discussions
with Peter Bartello, Raf Ferrari, Patrice Klein, Jim McWilliams, Jim
Riley, Shafer Smith, David Straub, Jacques Vanneste and Vladimir
Zeitlin are gratefully acknowledged. Comments and suggestions by three
reviewers helped improve both content and presentation.

\bibliography{/home/balu/Dropbox/BIB/ocean}{}

\end{document}